\documentclass[%
 aip,
 amsmath,amssymb,
preprint,%
]{revtex4-1}

\usepackage{graphicx}
\usepackage{dcolumn}
\usepackage{bm}

\usepackage[utf8]{inputenc}
\usepackage[T1]{fontenc}
\usepackage{mathptmx}

\usepackage{color}

\newcommand{\bd}{\begin{document}}
\newcommand{\ed}{\end{document}}
\newcommand{\bc}{\begin{center}}
\newcommand{\ec}{\end{center}}
\newcommand{\vs}{\vspace}
\newcommand{\hs}{\hspace}
\newcommand{\beq}{\begin{equation}}
\newcommand{\eeq}{\end{equation}}
\newcommand{\beqs}{\begin{eqn*}}
\newcommand{\eeqs}{\end{eqn*}}
\newcommand{\bq}{\begin{quote}}
\newcommand{\eq}{\end{quote}}
\newcommand{\lb}{\linebreak}
\newcommand{\mb}{\makebox}
\newcommand{\fb}{\framebox}
\newcommand{\mc}{\multicolumn}
\newcommand{\ben}{\begin{enumerate}}
\newcommand{\een}{\end{enumerate}}
\newcommand{\bit}{\begin{itemize}}
\newcommand{\eit}{\end{itemize}}
\newcommand{\ov}{\overline}
\newcommand{\un}{\underline}
\newcommand{\lt}{\left}
\newcommand{\rt}{\right}
\newcommand{\ba}{\begin{array}}
\newcommand{\ea}{\end{array}}
\newcommand{\beqa}{\begin{eqnarray}}
\newcommand{\eeqa}{\end{eqnarray}}
\newcommand{\beqas}{\begin{eqnarray*}}
\newcommand{\eeqas}{\end{eqnarray*}}
\newcommand{\bfg}{\begin{figure}}
\newcommand{\efg}{\end{figure}}
\newcommand{\pad}{\partial}
\newcommand{\nn}{\nonumber}
\newcommand{\la}{\leftarrow}
\newcommand{\ra}{\rightarrow}
\newcommand{\lgla}{\longleftarrow}
\newcommand{\lgra}{\longrightarrow}
\newcommand{\La}{\Leftarrow}
\newcommand{\Ra}{\Rightarrow}
\newcommand{\Lra}{\Leftrightarrow}
\newcommand{\Lgla}{\Longleftarrow}
\newcommand{\Lgra}{\Longrightarrow}
\renewcommand{\a}{\alpha}
\renewcommand{\b}{\beta}
\newcommand{\g}{\gamma}
\newcommand{\G}{\Gamma}
\renewcommand{\d}{\delta}
\newcommand{\D}{\Delta}
\newcommand{\e}{\epsilon}
\newcommand{\eps}{\epsilon}
\newcommand{\s}{\sigma}
\renewcommand{\l}{\lamda}
\newcommand{\m}{\mu}
\newcommand{\n}{\nu}
\renewcommand{\S}{\Sigma}
\newcommand{\p}{\pi}
\newcommand{\om}{\omega}
\newcommand{\Om}{\Omega}
\newcommand{\tri}{\triangle}
\newcommand{\ti}{\times}
\newcommand{\f}{\frac}
\newcommand{\ds}{\displaystyle}
\newcommand{\alter}[2]{\lt\{ \ba{ll}#1 \\ #2 \ea \rt.}
\newcommand{\altn}[4]{\lt\{ \ba{rl}#1 & \mb{if \, \,}#2 \\ #3 & \mb{}#4 \ea
    \rt.}
\newcommand{\altif}[4]{\lt\{ \ba{ll}#1 & \mb{if \, \,}#2 \\ #3 &
\mb{if \, \,}#4 \ea \rt.}
\newcommand{\altnif}[4]{\lt\{ \ba{rl}#1 & \mb{if \, \,}#2 \\ #3 &
\mb{if \, \,}#4 \ea \rt.}
\newcounter{algc}
\newcounter{romc}
\newcounter{Alphc}
\newcommand{\bl}{\begin{list}{{\it Step} ~\arabic{algc}~:} {\usecounter{algc}
                \setlength{\topsep}{0pt} \setlength{\itemsep}{0pt}}}
\newcommand{\el}{\end{list}}
\newcommand{\blr}{\begin{list}{~\roman{romc}~:} {\usecounter{romc}
                \setlength{\topsep}{0pt} \setlength{\itemsep}{0pt}}}
\newcommand{\elr}{\end{list}}
\newcommand{\bla}{\begin{list}{~\Alph{Alphc}~:} {\usecounter{Alphc}
                \setlength{\topsep}{0pt} \setlength{\itemsep}{0pt}}}
\newcommand{\ela}{\end{list}}
\newcommand{\tsup}{\textsuperscript}
\newcommand{\tsub}{\textsubscript}

\begin{document}


\title{Unified benchmarking and characterization protocol for nanomaterial-based heterogeneous photodetector technologies}

\author{Nithin Abraham}

\author{Kausik Majumdar}%
 \email{kausikm@iisc.ac.in}
 \affiliation{Department of Electrical Communication Engineering, Indian Institute of Science, Bangalore 560012, India.}


\begin{abstract}
Over the last few years, there has been a rapid growth towards demonstrating highly sensitive, fast photodetectors using photoactive nano-materials. As with any other developing and highly inter-disciplinary field, the existing reports exhibit a large spread in the data due to less optimized materials and non-standardized characterization protocols. This calls for a streamlined performance benchmarking requirement to accelerate technological adoption of the promising candidates. The goal of this paper is four-fold: (i) to address the key challenges to perform such benchmarking exercise; (ii) to elucidate the right figures of merit to look for, and in particular, to demonstrate that noise-equivalent-power ($N\!E\!P$) is a more reliable sensitivity metric than other commonly used ones, such as responsivity ($R$) and specific detectivity ($D^*$); (iii) to propose $N\!E\!P$ versus frequency of operation ($f$) as a single, unified benchmarking chart that could be used for apple-to-apple comparison among heterogeneous detector technologies; and (iv) to propose a simple, yet streamlined characterization protocol that can be followed while reporting photodetector performance.
\end{abstract}

\maketitle

\section*{Introduction}\label{sec:intro}
The advent of nano-materials not only brings an unprecedented opportunity to explore novel scientific phenomena at the nanoscale, they are of practical interest in various real life device applications. Nanomaterials have demonstrated significant performance improvement and remarkable new properties in electronic and optoelectronic devices. Photodetector is one such example where researchers have extensively explored the use of different nano-materials with varying dimensionality, including quantum dots, nanowires, two-dimensional layered materials, plasmonic materials, persovskites and a heterojunction of these in a variety of novel device structures \cite{Konstantatos2010,Zhai2010,Koppens2014,Saran2016,Fang2017,Long2018}. The intense research in this topic in the recent years shows promise to provide low cost solutions in different applications including remote sensing, medical imaging, consumer electronics, public safety, space, military and industrial instrumentation.

On the flip side, the growth technique of the relatively new nano-materials are often less optimized and lack in comprehensive characterization, for example, in terms of defect density \cite{Rana2017,Lou2013} and their energy distribution - which play a profound role in the photodetection characteristics. In addition, being an interdisciplinary area, researchers contributing from different fields often use different notions of performance metrics. The situation is aggravated by the lack of a standard measurement protocol, which has led to several claims of "record high" performance detectors, while in reality, many of these detectors actually perform rather poorly. Photodetectors should therefore be designed with a clarity about the appropriate figures of merit to target to realize the full potential of the nanomaterials. It is thus of utmost importance (1) to formulate a unified benchmarking methodology where detectors from heterogeneous technologies can be compared in a facile and seamless manner, and (2) to build a strict and standardized, yet simple characterization protocol that research groups can follow while reporting the performance of a photodetector. This article elucidates on both these aspects.

\section*{Nanomaterials for photodetection}\label{sec:why}
Nanomaterials provide unique advantages over their bulk counterpart in terms of photodetection, and thus nanomaterial based photodetectors have seen an unparalleled growth in recent years. Researchers have come up with different device designs to exploit the low dimensionality of these materials to their benefit. Despite the suppressed light absorption due to reduced thickness of material, nanomaterials are advancing as preferred candidates for photodetection because of the exciting possibilities they offer. Concerning photodetection, one of the highly desired features is the ability to tune the absorption spectrum of the material. By controlling the growth and in turn the features of the nanomaterial, for example the size of a quantum dot or shape of a nanoparticle, one can modify the absorption spectrum of the device \cite{Norris1996,Wu1987,Kongkanand2008}. By doing so, one can choose to tune the wavelength range of absorption, and to reject a particular band. Apart from growth parameters, strain can also be introduced for the same \cite{Rahneshin2016,Lee2017}. The absorption can also be tuned by designing suitable plasmonic nanostructures and nanoantennas \cite{Miao2019,Lin2019}. The low dimensionality of these devices also opens new possibilities. For a photodetector using photovoltaic effect as the detection mechanism, the efficiency depends on how efficiently the built-in field can separate the carriers. This field can be several orders of magnitude higher in a vertical nano-film based detector compared to conventional bulk detectors\cite{Zhou2018,Long2016}. Another interesting aspect is the ability to form atomically sharp heterostructures, especially with two-dimensional layered materials, which are otherwise unattainable in bulk semiconductor junctions. These ultra-thin, near-lattice-matched layered heterostructures act as a new material by themselves and exhibit a wide range of fascinating electrical and optical phenomena. Extremely fast photodetection is achieved by the ultrafast inter-layer charge separation in these vertical layered heterostructures \cite{Koppens2015,Zhang2017}. Ultra-fast photodetection has also been demonstrated using the high mobility of graphene - the thinnest known material \cite{Xia2009}. Another unique advantage of nanomaterial based photodetectors is that they provide ease of gating, and thus the obtained photocurrent can be gate controllable \cite{Konstantatos2015,Kis2013}. The ultra-thin nature of the active material also helps to suppress the dark current very significantly, which in turn results in a significantly reduced noise and hence enhanced sensitivity. These advantages, coupled with the low cost of fabrication, make nanomaterials very promising for highly sensitive photodetection applications.

The working principle of different nanomaterial based photodetectors can be categorized into one or more of the following effects:

\textbf{a) Photoconductive effect:} Photoconductivity is the result of a change in electrical conductivity of the material due to the excess optically generated carriers. Photoconductive devices often offer a large gain. This arises due to successive re-injection of one type of carrier (electron or hole) to maintain charge neutrality when the other type of carrier remains in the channel, either due to lower mobility or due to trapping\cite{Li2019}.

\textbf{b) Photogating effect:} In photogating, one type of the optically generated carriers gets trapped in another layer or trap states which can then electrostatically gate the channel. This method is usually associated with a large gain\cite{Luo20182,Zhang2019}. Depending on the channel type and the type of the trapped carriers, this shifts the transfer curve of the device to the left or to the right leading to an increase or decrease in the channel current. Depending on the lifetime of the trapped carriers responsible for gating, the speed of the photodetector varies.

\textbf{c) Photovoltaic effect:} Photovoltaic effect is observed in devices with a built-in field, for example, in p-i-n diodes. The built-in field causes photogenerated electron and hole to move in opposite directions and would then be collected at different contacts\cite{Liu2019,Liu2018}. The photoresponse is usually very fast in photovoltaic devices.

\textbf{d) Photo-thermoelectric Effect:} Photo-thermoelectric effect happens in devices where the light illumination causes a temperature gradient along the current path. The temperature gradient along with a difference in the Seebeck coefficients creates a detectable photo-voltage\cite{Viti2019}.

\textbf{e) Photo-bolometric Effect:} Photo-bolometric effect is the change in the resistivity of a material due to light induced heating. The sign of the photocurrent depends on the sign of the temperature coefficient of the active material\cite{Ryzhii2019,Liu20192}. The speed of the detector depends on how fast the material can restore its initial state.

The performance and noise characteristics of the photodetectors employing the different effects vary significantly, and pose significant challenge to benchmarking. For example, in a device utilizing photogating from trapped carriers, the random fluctuations in the trapped carrier density causes fluctuations in the shift in transfer curve. This will in turn show up in the device current. Depending on the gain of the device, this noise can be significantly amplified. In the above example, the trap density will largely depend on the growth technique and parameters employed in fabrication of the device. Another major consideration is the low dimensionality of nanomaterials, which often makes the photocurrent cross-section and photo-active area of the devices different. For example, in case of a nanowire based photodetector, these two areas are typically two to three orders of magnitude different\cite{Sett2018,Butanovs2018,Luo2018,Chen2018}. Usual benchmarking metric such as specific detectivity would give rise to unrealistic values and completely fails under such ultra-high aspect ratio scenarios. Conventional methods of benchmarking photodetectors thus must be revisited for nanomaterial based devices.

\section*{\label{sec:level2}Benchmarking challenges}
\begin{figure*}[!t]
\centering
\includegraphics[scale=0.85]{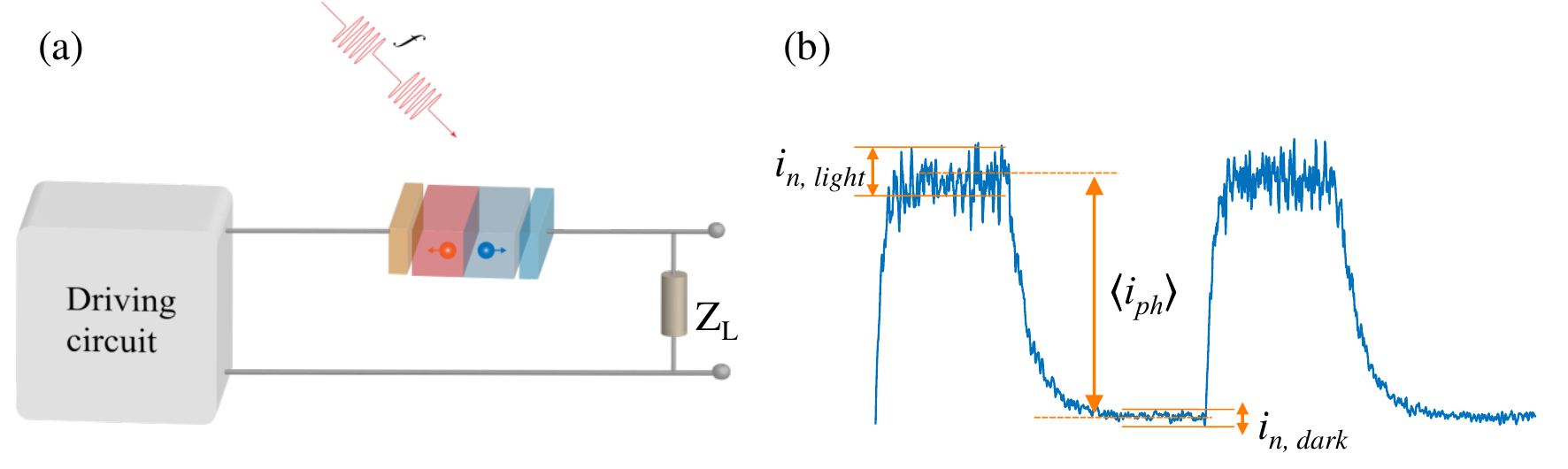}
\vspace{-0.2in}
\caption{\label{fig:setup}\textbf{Typical photodetection setup.} (a) A schematic view depicting a basic photodetection measurement setup. $f$ is the modulation frequency of the incident light and $Z_L$ is the load across which output is measured. (b) Typical transient response of a photodetector, indicating average photocurrent ($\langle i_{ph} \rangle$), r.m.s. noise current under dark condition and in presence of light.}
\end{figure*}

The working principle of a typical photodetector is explained schematically in Figure \ref{fig:setup}. The performance of the detector can be judged by four parameters: (1) The wavelength range in which the photodetector works, (2) the sensitivity of the detector, that is how weak an incoming signal can be reliably detected by the photodetector, (3) the dynamic range of the detector, and (4) how fast the photodetector is able to respond to an incoming signal. A benchmarking chart allows one to compare the performance among different devices in a technology domain, and is extremely useful for the steady development of that technology. Unfortunately, the way the nano-material based photodetectors are characterized and the data is presented in the current literature, there are several challenges that one must address in order to build a meaningful benchmarking chart where the ``true" performance can be compared among different reported detectors in an apple-to-apple manner.
\\\\
\textbf{(a) Lack of linear dynamic range:} Linear dynamic range ($L\!D\!R$) indicates the range of the input optical power where the detector output signal remains a linear function of it. In a majority of the nano-material-based photodetectors, in particular, those which report high gain, $L\!D\!R$ is obtained only at extremely low optical power level (sometimes in tens of fW). The $L\!D\!R$ is thus often missed in detectors where the detector noise does not allow detection of such low input power levels. Such a deviation from linearity with an increase in the optical power is understood as a combination of one or more of these effects: (1) saturation of filling of trap states by photocarriers, (2) enhanced recombination of photogenerated electron-hole pairs, (3) suppression of electric field due to increased screening arising from enhanced photocarrier density, and (4) relatively higher series resistance at higher optical power. The photocurrent in the $L\!D\!R$ regime is frequently extrapolated to obtain the minimum detectable power by a photodetector \cite{Fang2019}. In the situation where the photodetector does not exhibit $L\!D\!R$, one must be careful while performing such extrapolation. The variation of performance with incident optical power poses a substantial challenge in choosing the optical power while benchmarking different photodetectors.
\\\\
\textbf{(b) Need for an accurate noise characterization:} The noise of the detector limits the least possible power that can be detected by the device and thus limits the sensitivity of the detector. The estimation of noise remains one of the most neglected aspects of performance metric in nanomaterial-based detectors reported in recent times \cite{Fang2019}. The noise in photodetectors originates from a wide array of sources like photoinduced noise (g-r noise), transport noise, contact noise, and modulated electronic noise, many of which are not necessarily white in nature \cite{vanVliet67}. Disregarding these, the detector noise is most frequently modeled as a white noise, assuming it arises from the fluctuations in the dark current and the thermal Johnson noise. More often than not, the Johnson noise and the flicker ($1/f$) noise are neglected, and only the shot noise component coming from the dark current is considered. While the above approach is attractive due to its simplicity, it may predict a substantially lower $N\!E\!P$ from the actual value, particularly in detectors with an internal gain. It would be a quite formidable task to have a realistic model of the noise behaviour of an arbitrary device geometry, with an active nano-material of unknown trap distribution or a detector employing a novel detection mechanism. A more reliable way to estimate the noise characteristics is to measure the noise power spectrum of the detector, as explained later. Clearly, lack of appropriate noise data of reported photodetectors poses a significant challenge to true performance comparison.

\begin{figure*}[!t]
\centering
\includegraphics[scale=0.85]{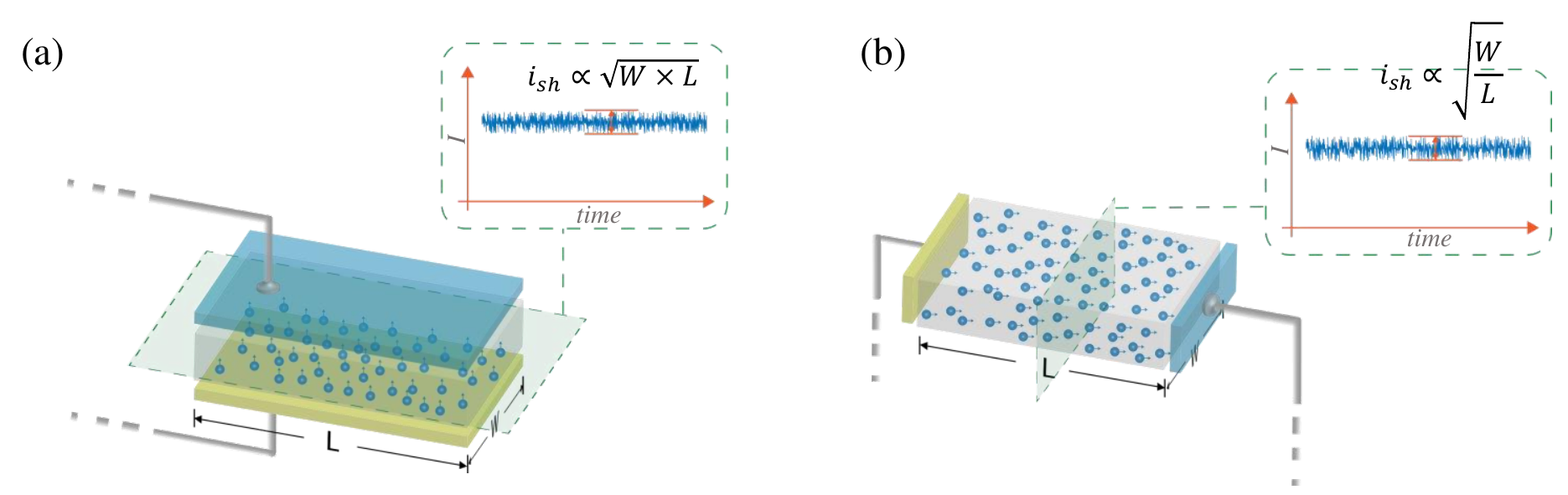}
\vspace{-0.2in}
\caption{\label{fig:noise}\textbf{Area dependence of shot noise.} Illustration of carrier flow in vertical (a) and lateral (b) devices indicating different area scaling of shot noise, making $N\!E\!P$ a more reliable figure of merit for sensitivity compared with $D^*$.}
\end{figure*}

\textbf{(c) $N\!E\!P$ is a more generalized sensitivity metric than $D^*$:} The sensitivity of a photodetector is determined by the signal-to-noise ratio ($S\!N\!R$), which is often expressed in terms of the noise equivalent power ($N\!E\!P$). The responsivity $R$ of a detector is defined as
\beq\label{eq:R}
R = \frac{\langle i_{light}\rangle - \langle i_{dark}\rangle}{P_{op}} = \frac{\langle i_{ph} \rangle}{P_{op}}
\eeq
where the angular brackets represent the average values of the current with light ($i_{light}$) and without light ($i_{dark}$), $i_{ph} = i_{light} - i_{dark}$ is the photocurrent (see Figure \ref{fig:setup}b), and $P_{op}$ is the incident optical power. If $i_n$ is the root mean squared (r.m.s.) noise current (typically band limited to $1$ Hz, that is, with an integration time of $0.5$ s \cite{ThorlabsNEP}) from the detector, we have
\beq\label{eq:i_n_by_R}
\frac{i_n}{R} = P_{op}\frac{i_n}{\langle i_{ph} \rangle} = \frac{P_{op}}{S\!N\!R}
\eeq
$N\!E\!P$ is then defined as
\beq\label{eq:NEP}
N\!E\!P = \frac{i_n}{R}\Big|_{S\!N\!R=1}
\eeq

\begin{figure*}[!t]
\centering
\includegraphics[scale=0.85]{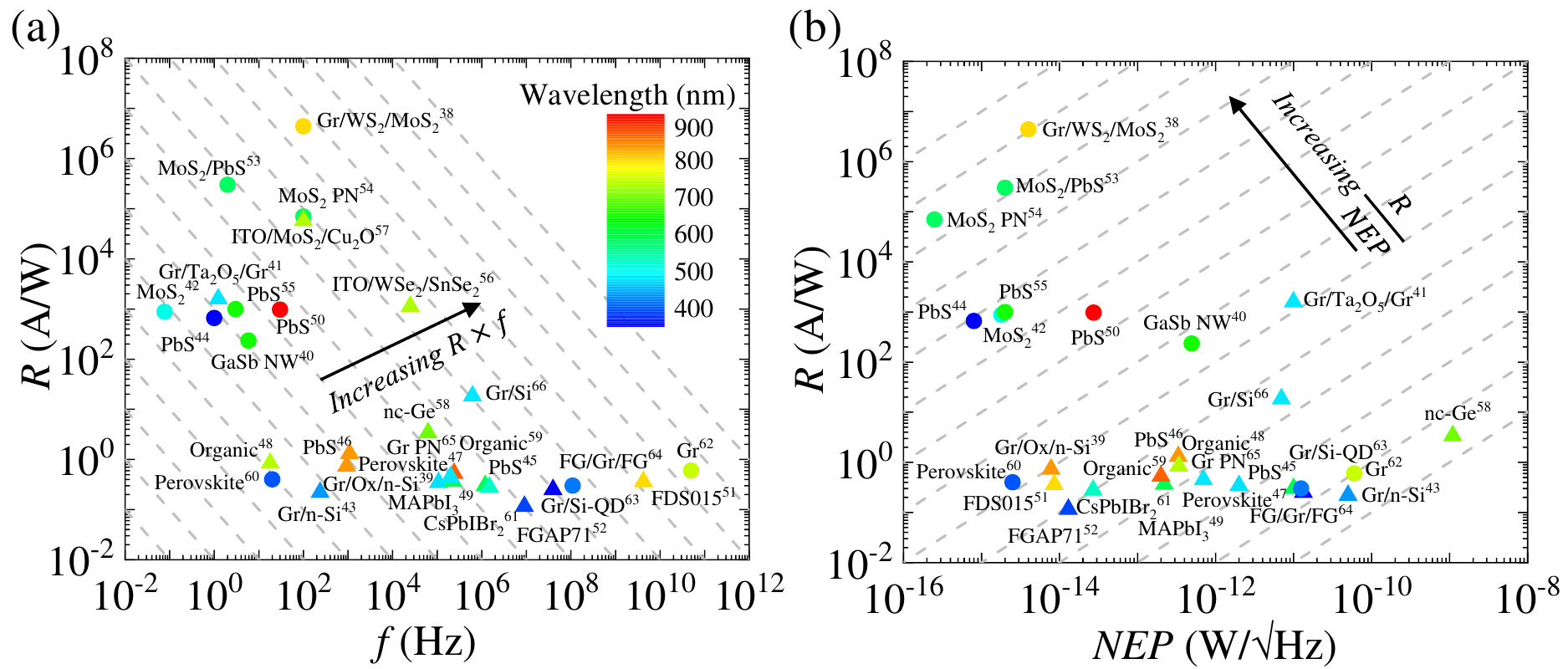}
\caption{\label{fig:R_NEP_f}\textbf{Benchmarking photodetectors.} (a) Responsivity ($R$) versus modulation frequency ($f$) and (b) $R$ versus $N\!E\!P$ of various nano-material based photodetectors\cite{Murali2019,Li2016,Luo2015,Liu2014,Lopez-Sanchez2013,An2013,Konstantatos2007,Kim2015,Shen2016,Casaluci2016,Wang2018,Sutherland2015,Konstantatos2006,FDS015,FGAP71,Kufer2015,Huo2017,Hwang2016,Murali2018,Kallat2018,Dhyani20,Huang20,Li20,Bao18,Cakmakyapan2018,Yu16,Xu17,Kim2014,Huang18}. The color of individual data point represents the wavelength of operation, as indicated by the color bar in the inset of (a). The triangles (circles) in (a) and (b) represent devices employing vertical (lateral) current transport. The parallel dashed lines indicate loci of constant gain-bandwidth product in (a) and constant $R$-$N\!E\!P$ ratio in (b). All data points correspond to room temperature measurement.}
\end{figure*}

Hence, $N\!E\!P$ is the input signal power that results in a unity $S\!N\!R$ at a $1$ Hz output bandwidth. If $i_n$ is measured with a bandwidth limitation of $BW$, the result is further divided by $\sqrt{BW}$ in order to normalize the obtained $N\!E\!P$ to $1$ Hz. To achieve high sensitivity, one thus aims to maximize the responsivity and minimize the detector noise (which depends on both dark noise as well as the noise in presence of light).
\\\\
In this context, specific detectivity ($D^*$) is a popular figure of merit to measure the sensitivity of detectors, and is defined as
\beq\label{eq:D*}
D^* = \frac{\sqrt{A}}{N\!E\!P}
\eeq
where $A$ is the active device area. $D^*$ is useful to compare sensitivity of detectors with varying active area. However, one must remember that such a comparison only makes sense when the noise scales as $\sqrt{A}$, such as shot noise. For example, using this, one can compare two vertical detectors where the source of noise is primarily governed by the dark current, which in turn scales with the active area. However, as discussed above, there could be a variety of noise sources for different detectors, based on the geometry and the active nano-material, which can dominate over shot noise.
\\\\
Even when the detector noise is dominated by shot noise, the area scaling of the noise may not be present, as explained using a simple example in Figure \ref{fig:noise}. In a vertical detector where photocarriers are transported vertically (Figure \ref{fig:noise}a), the dark shot noise is proportional to $\sqrt{A} = \sqrt{W\times L}$, and use of $D^*$ is justified. On the other hand, in a planar detector where photocarriers are transported laterally (Figure \ref{fig:noise}b), the dark shot noise is proportional to $\sqrt{\frac{W}{L}}$, and not to $\sqrt{A}$. Thus scaling the noise to $\sqrt{A}$ results in an unreliable value for $D^*$. In particular, the noise current in the latter device, for a square active area, is completely independent of $W$ and $L$. So it is imperative to note that $D^*$ is valid only for devices where shot noise is the dominating component of the noise, and the optically active area and current flow cross-section area are the same. In several of recently reported photodetectors, including the planar ones, these assumptions do not hold good, making $D^*$ an unreliable metric. Care should thus be taken while reporting $D^*$ of a photodetector. On the other hand, $N\!E\!P$ does not have any such constraint and we propose to use $N\!E\!P$ (or, $S\!N\!R$) as a more generalized measure of sensitivity.
\\\\

\begin{figure}[!t]
\centering
\includegraphics[scale=0.5]{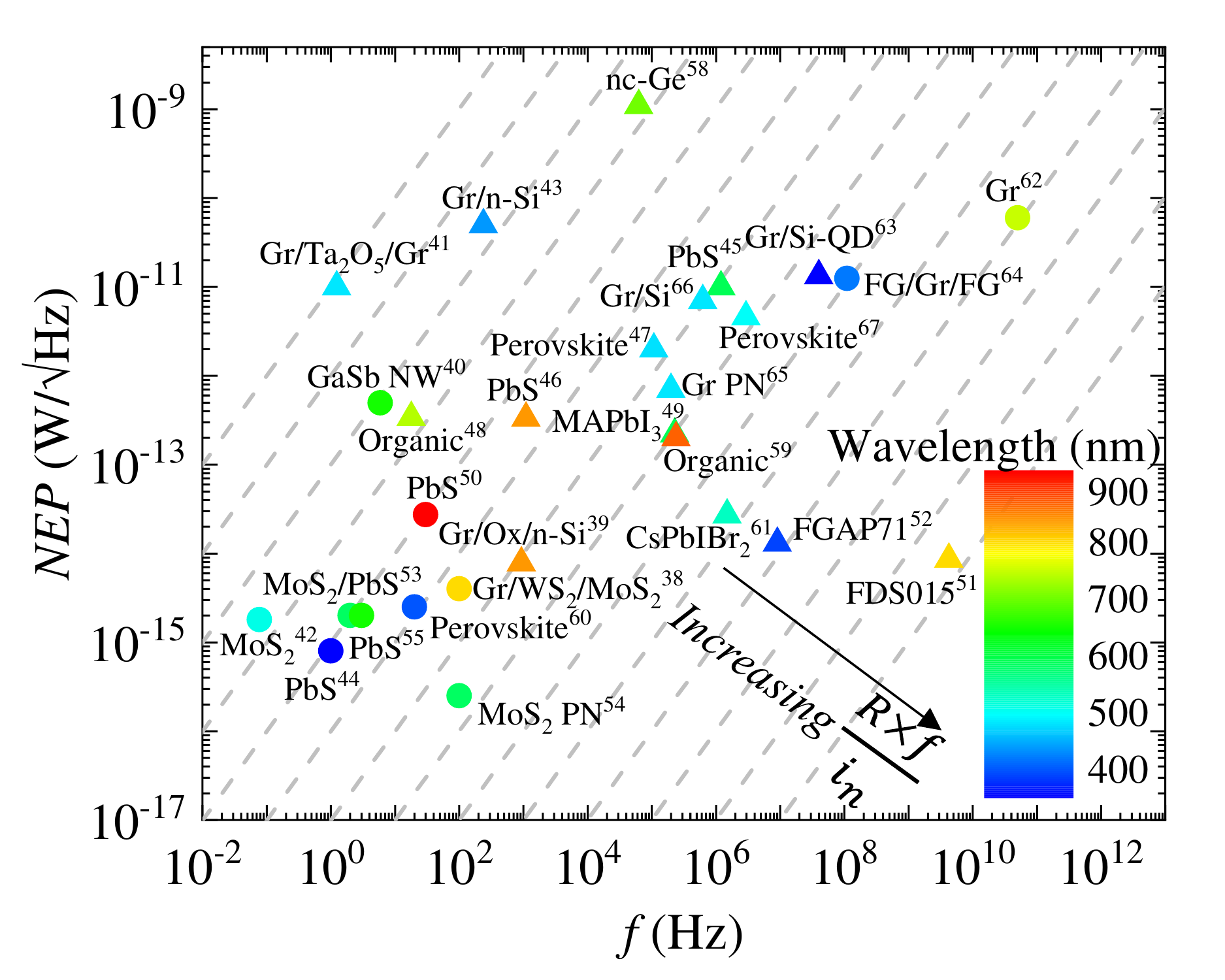}
\caption{\label{fig:NEP_f}\textbf{Proposed benchmarking chart - $N\!E\!P$ versus $f$.} $N\!E\!P$ versus modulation frequency ($f$) of various nano-material based photodetectors\cite{Murali2019,Li2016,Luo2015,Liu2014,Lopez-Sanchez2013,An2013,Konstantatos2007,Kim2015,Shen2016,Casaluci2016,Wang2018,Sutherland2015,Konstantatos2006,FDS015,FGAP71,Kufer2015,Huo2017,Hwang2016,Dou2014,Dhyani20,Huang20,Li20,Bao18,Cakmakyapan2018,Yu16,Xu17,Kim2014,Huang18}. The triangles (circles) represent devices employing vertical (lateral) current transport. The parallel dashed lines indicate loci of constant frequency-$N\!E\!P$ ratio. A higher value for this ratio indicates a better overall device performance.}
\end{figure}

\textbf{(d) Responsivity versus speed of operation:} The photodetection mechanism in nano-material based detectors is often dominated by traps, which usually capture one type of carriers and result in a large internal gain due to photoconductive and photogating effects. Since trapping and de-trapping of charge carriers from trap sites are relatively slow phenomena, the detectors that overly rely on such gain mechanism tend to exhibit a large response time \cite{Konstantatos2018}. At the extreme condition, it often takes a very long time (often on the order of several seconds to minutes) to de-trap charge carriers from trap sites which are deep inside the bandgap, resulting in a large reminiscent photocurrent. In order to establish the point, we plot $R$ values collected from literature as a function of modulation frequency ($f$) in Figure \ref{fig:R_NEP_f}a. Since most of the articles only report the response time, rather than the bandwidth, wherever bandwidth information is not available, we take $f = \frac{1}{t_r + t_f}$ where $t_r$ and $t_f$ are the $10\%$-$90\%$ rise and fall times, respectively. In the same plot, we also indicate the loci of constant $R\times f$ which represent constant gain-bandwidth product lines. There is a clear trend of suppressed speed of operation as the reported $R$ increases, which is not desirable in many applications. In this context, segregating the data from Figure \ref{fig:R_NEP_f}a based on the current transport directions indicates that devices employing vertical transport (triangle symbols) largely outperform those with lateral transport (circle symbols) in terms of operating frequency (with the exception of graphene based detectors due to its high in-plane mobility). This is mainly due to the ultra-short transient time for vertical transport in majority of these systems. However, employing gain mechanisms like photo-gating becomes difficult compared to lateral devices, which leads to a weaker responsivity in vertical devices. Nonethless, some intelligent device designs has been reported in the literature circumventing these difficulties. We should mention two important points in this context. First, in several reports, the measurement conditions for $R$ and $f$ are different. For the detectors that are not working in the linear dynamic range, $R$ is typically reported at the minimum detectable input power to maximize the reported $R$ value, while $f$ is measured at higher power levels - adding to the confusion. Second, the measurement conditions of $R$ vary significantly from one report to another, for example, in terms of applied bias and incident optical power, making it difficult to compare the reported results.

\textbf{(e) High responsivity does not necessarily mean high sensitivity:} In a photodetector, while it is desirable to have a high $R$, in the recent literature, there has been a significant degree of over emphasis on $R$ as a performance metric. What is often ignored in high gain detectors is the large gain noise present in these devices. The point is explained in Figure \ref{fig:R_NEP_f}b by plotting $R$ as a function of $N\!E\!P$ from recently reported detectors. Among these, most of the $N\!E\!P$ values are based on dark noise measurement. What is surprising in Figure \ref{fig:R_NEP_f}b is that the $N\!E\!P$ shows very low correlation with $R$, and does not corroborate well with Equation \ref{eq:NEP}. This points to a large variation in the noise current among reported detectors, and clearly shows that an increase in $R$ does not necessarily indicate a high sensitivity. As a matter of fact, one can in principle achieve a high sensitivity with a relatively low responsivity (bottom-left corner of the chart) as long as the noise of the detector is kept under control. The trend in literature to maximize $R$, without worrying about the response time and the noise performance can lead to a poor photodetector design.
\section*{\label{sec:level8}Proposed unified benchmarking and characterization protocol}
In the light of the above discussion, it is desirable to have a single unified benchmarking chart that can be used to compare the performance of different photodetectors. A benchmarking chart consisting of $N\!E\!P$ versus $f$ provides a direct measure of the usefulness of photodetection devices in terms of sensitivity (which embeds both responsivity and noise) and speed of operation. One such chart is constructed from different data points available in Figure \ref{fig:NEP_f}. The chart suggests that the operational speed is often compromised in order to achieve low $N\!E\!P$. Also, the trend of lateral transport photodetectors (circle symbols) occupying the low $N\!E\!P$ regions of the chart and the vertical photodetectors (triangle symbols), the high $N\!E\!P$ regions emphasize that the noise current in these devices have significant contributions from sources which scale with the current cross-section area and not the photo active area of the device. This supports our reasoning of the universality of $N\!E\!P$ over $D^*$. It is worth noting that the commercial detectors \cite{FDS015,FGAP71} still largely outperform the research grade detectors, particularly in terms of speed of operation, while maintaining a high degree of sensitivity. This suggests that a large window of opportunity exists for nano-material based photodetection schemes.
\\
\begin{figure*}[!t]
\centering
\includegraphics[scale=0.55]{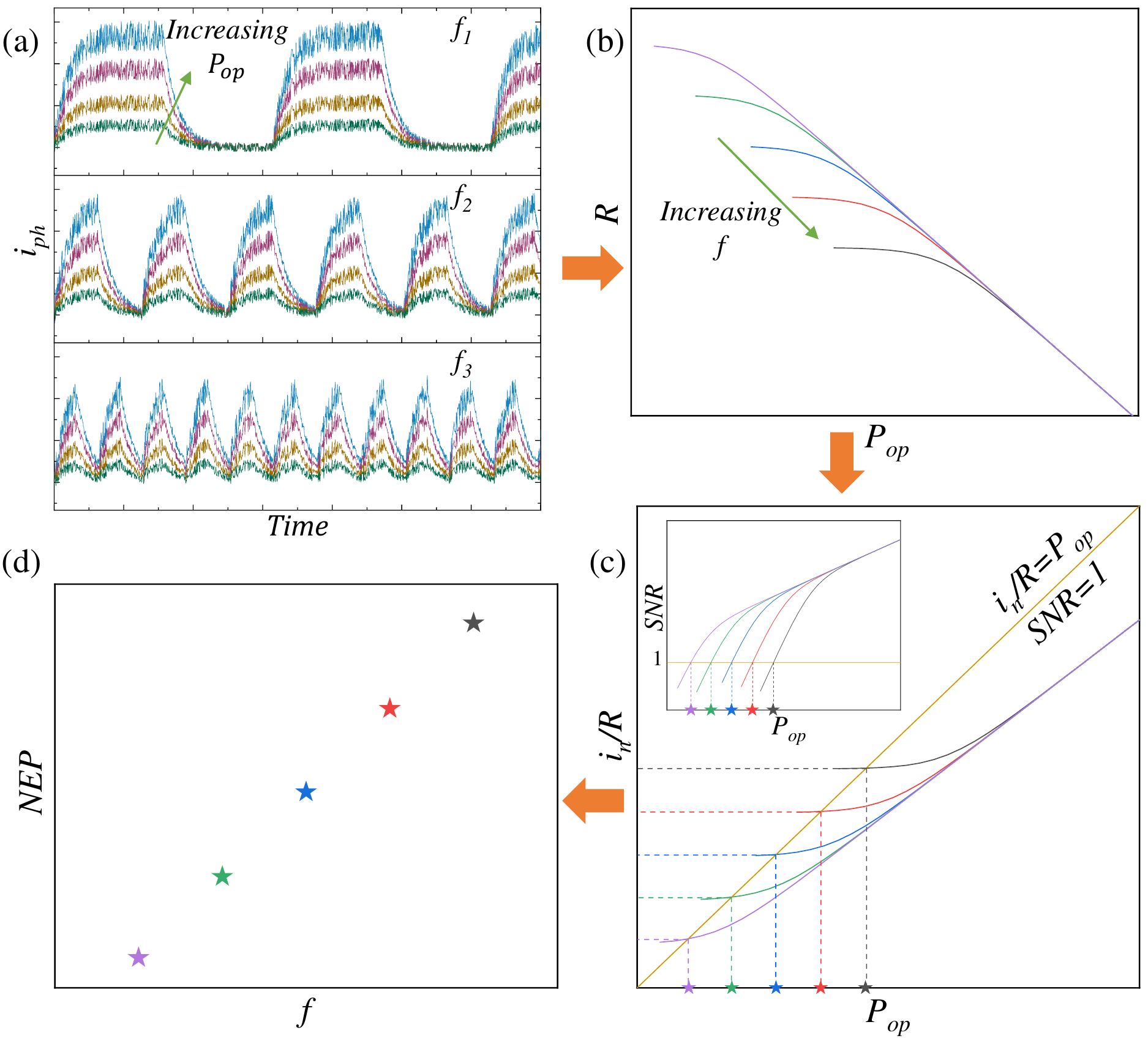}
\caption{\label{fig:protocol}\textbf{Proposed characterization protocol.} (a) Typical transient response of a photodetector under various illumination intensities $P_{op}$ and modulation frequencies ($f$), with $f_1<f_2<f_3$. (b) Extracted Responsivity as a function of $P_{op}$ at various modulation frequencies. (c) Ratio of measured r.m.s. noise current to responsivity as a function of $P_{op}$ for different $f$. Orange line corresponds to $S\!N\!R = 1$, intersection of which with the measured curves gives $N\!E\!P$. Inset: $S\!N\!R$ of the detector versus $P_{op}$ for different $f$. The $N\!E\!P$ value is obtained at $S\!N\!R = 1$. (d) The $N\!E\!P$ obtained as a function of $f$ which is used in the final benchmarking chart to compare with other detectors.}
\end{figure*}

Keeping the above discussions in mind, we now propose a performance evaluation and benchmarking methodology in a step by step manner. The proposed methodology is based on a time domain measurement. The methodology will help to evaluate the performance of the detectors from heterogeneous technologies in a unified manner, and will allow a fair comparison among different competing detectors.
\\
\textbf{Step 1 - Choose the wavelength range:} Detectors operating at different wavelength ranges exhibit significantly different properties, and typically a different set of challenges are aimed at. In order to have an apple to apple comparison, the first step is to identify the desired range of wavelength of operation of the detector, typically governed by the user specification.
\\
\textbf{Step 2 - Measure $i(P_{op},f,t)$:} The next step is to measure the transient response of the detector. This is done by modulating the intensity of a well-calibrated light source at different frequencies $f$ and at different incident optical power $P_{op}$ and recording the detector output. A typical transient response of a detector as a function of $P_{op}$ and $f$ is shown in Figure \ref{fig:protocol}a.
\\
\textbf{Step 3 - Extract $R(P_{op},f)$:} Next the responsivity $R(P_{op},f)$ of the detector is extracted as a function of $P_{op}$ and $f$ using Equation \ref{eq:R}, as explained in Figure \ref{fig:protocol}b. Ideally, for a given frequency of operation, $P_{op}$ should be gradually reduced until the linear dynamic range is achieved, that is, where $R$ becomes independent of $P_{op}$. However, depending on the noise characteristics of the detector, it may not be possible to reach this regime of operation. For detector with high gain, as the frequency of modulation is increased, the gain of the detector may go down and hence the responsivity, as shown in Figure \ref{fig:protocol}b.
\\
\textbf{Step 4 - Extract $N\!E\!P$:} The next step is to find out the noise current $i_n$ and $N\!E\!P$. As mentioned earlier, the spectral distribution of the detector noise is a strong function of the material properties and the design of the individual detector. In order to standardize the portion of the noise spectrum taken to calculate $N\!E\!P$, $i_n$ is measured limited to $1$ Hz output bandwidth. It is important to characterize the noise under illumination and dark as the gain noise of the detector can be high depending on the specifics of the detector and can influence the SNR under illumination. The variation of NEP obtained by considering the noise in the presence of light to that under dark condition gives an idea of the extent of gain noise contribution to the device noise spectrum. If there is no significant deviation, NEP obtained using noise characterization under dark condition is justified. The quantity $\frac{i_n}{R} (= \frac{P_{op}}{S\!N\!R})$ is then plotted against $P_{op}$ for different $f$, as shown in Figure \ref{fig:protocol}c. In the same plot, we also show the $S\!N\!R=1$, that is $\frac{i_n}{R} = P_{op}$ line. The values of the $P_{op}$ at the intersection points at different frequencies of operation represent the $N\!E\!P$ of the detector. Alternatively, the same can be achieved by plotting the $S\!N\!R$ as well. In the inset of Figure \ref{fig:protocol}c, $S\!N\!R$ is plotted as a function of $P_{op}$, and the values of $P_{op}$ at $S\!N\!R=1$ are taken as the $N\!E\!P$ at the respective frequencies. Note that no noise modeling is required in this procedure avoiding unrealistic assumptions. The measured noise takes into account all possible sources of the noise, and will typically be larger than the dark shot noise.
\\
\textbf{Step 5 - Plot $N\!E\!P$ versus modulation frequency $f$:} Finally, in order to create the chart for figure of merit to benchmark against other detectors, the obtained $N\!E\!P$ values at different modulation frequencies $f$ are plotted as a function of $f$, as shown in Figure \ref{fig:protocol}d.
This single chart summarizes the information about the detector performance in terms of sensitivity and speed of operation and can be constructed for any detector irrespective of the active material, geometry, and photodetection mechanism.
\\\\
In summary, a streamlined characterization protocol and careful benchmarking exercise, as described above, will help to compare the true performance of different detectors, and thus isolate the promising photodetector candidates from the rest of the lot, which is essential for rapid industrial adoption. The findings will also help to steer the research efforts of novel detectors in a fruitful direction.
\section*{Acknowledgements}
The authors thank Prof. Arindam Ghosh and Dr. Sumesh M. A. for useful discussions. This work was supported in part by a grant under Indian Space Research Organization (ISRO), by the grants under Ramanujan Fellowship, Early Career Award, and Nano Mission from the Department of Science and Technology (DST), and by a grant from MHRD, MeitY and DST Nano Mission through NNetRA.
\section*{Competing Interests}
The Authors declare no Competing Financial or Non-Financial Interests.
\section*{Data availability}
	Data available on request from the authors.
\bibliography{ref}

\begin{thebibliography}{68}%
\makeatletter
\providecommand \@ifxundefined [1]{%
 \@ifx{#1\undefined}
}%
\providecommand \@ifnum [1]{%
 \ifnum #1\expandafter \@firstoftwo
 \else \expandafter \@secondoftwo
 \fi
}%
\providecommand \@ifx [1]{%
 \ifx #1\expandafter \@firstoftwo
 \else \expandafter \@secondoftwo
 \fi
}%
\providecommand \natexlab [1]{#1}%
\providecommand \enquote  [1]{``#1''}%
\providecommand \bibnamefont  [1]{#1}%
\providecommand \bibfnamefont [1]{#1}%
\providecommand \citenamefont [1]{#1}%
\providecommand \href@noop [0]{\@secondoftwo}%
\providecommand \href [0]{\begingroup \@sanitize@url \@href}%
\providecommand \@href[1]{\@@startlink{#1}\@@href}%
\providecommand \@@href[1]{\endgroup#1\@@endlink}%
\providecommand \@sanitize@url [0]{\catcode `\\12\catcode `\$12\catcode
  `\&12\catcode `\#12\catcode `\^12\catcode `\_12\catcode `\%12\relax}%
\providecommand \@@startlink[1]{}%
\providecommand \@@endlink[0]{}%
\providecommand \url  [0]{\begingroup\@sanitize@url \@url }%
\providecommand \@url [1]{\endgroup\@href {#1}{\urlprefix }}%
\providecommand \urlprefix  [0]{URL }%
\providecommand \Eprint [0]{\href }%
\providecommand \doibase [0]{http://dx.doi.org/}%
\providecommand \selectlanguage [0]{\@gobble}%
\providecommand \bibinfo  [0]{\@secondoftwo}%
\providecommand \bibfield  [0]{\@secondoftwo}%
\providecommand \translation [1]{[#1]}%
\providecommand \BibitemOpen [0]{}%
\providecommand \bibitemStop [0]{}%
\providecommand \bibitemNoStop [0]{.\EOS\space}%
\providecommand \EOS [0]{\spacefactor3000\relax}%
\providecommand \BibitemShut  [1]{\csname bibitem#1\endcsname}%
\let\auto@bib@innerbib\@empty
\bibitem [{\citenamefont {Konstantatos}\ and\ \citenamefont
  {Sargent}(2010)}]{Konstantatos2010}%
  \BibitemOpen
  \bibfield  {author} {\bibinfo {author} {\bibfnamefont {G.}~\bibnamefont
  {Konstantatos}}\ and\ \bibinfo {author} {\bibfnamefont {E.~H.}\ \bibnamefont
  {Sargent}},\ }\bibfield  {title} {\enquote {\bibinfo {title} {Nanostructured
  materials for photon detection},}\ }\href {\doibase 10.1038/nnano.2010.78}
  {\bibfield  {journal} {\bibinfo  {journal} {Nature Nanotechnology}\ }\textbf
  {\bibinfo {volume} {5}},\ \bibinfo {pages} {391 EP --} (\bibinfo {year}
  {2010})}\BibitemShut {NoStop}%
\bibitem [{\citenamefont {Zhai}\ \emph {et~al.}(2010)\citenamefont {Zhai},
  \citenamefont {Li}, \citenamefont {Wang}, \citenamefont {Fang}, \citenamefont
  {Bando},\ and\ \citenamefont {Golberg}}]{Zhai2010}%
  \BibitemOpen
  \bibfield  {author} {\bibinfo {author} {\bibfnamefont {T.}~\bibnamefont
  {Zhai}}, \bibinfo {author} {\bibfnamefont {L.}~\bibnamefont {Li}}, \bibinfo
  {author} {\bibfnamefont {X.}~\bibnamefont {Wang}}, \bibinfo {author}
  {\bibfnamefont {X.}~\bibnamefont {Fang}}, \bibinfo {author} {\bibfnamefont
  {Y.}~\bibnamefont {Bando}}, \ and\ \bibinfo {author} {\bibfnamefont
  {D.}~\bibnamefont {Golberg}},\ }\bibfield  {title} {\enquote {\bibinfo
  {title} {Recent developments in one-dimensional inorganic nanostructures for
  photodetectors},}\ }\href {\doibase 10.1002/adfm.201001259} {\bibfield
  {journal} {\bibinfo  {journal} {Advanced Functional Materials}\ }\textbf
  {\bibinfo {volume} {20}},\ \bibinfo {pages} {4233--4248} (\bibinfo {year}
  {2010})}\BibitemShut {NoStop}%
\bibitem [{\citenamefont {Koppens}\ \emph {et~al.}(2014)\citenamefont
  {Koppens}, \citenamefont {Mueller}, \citenamefont {Avouris}, \citenamefont
  {Ferrari}, \citenamefont {Vitiello},\ and\ \citenamefont
  {Polini}}]{Koppens2014}%
  \BibitemOpen
  \bibfield  {author} {\bibinfo {author} {\bibfnamefont {F.~H.~L.}\
  \bibnamefont {Koppens}}, \bibinfo {author} {\bibfnamefont {T.}~\bibnamefont
  {Mueller}}, \bibinfo {author} {\bibfnamefont {P.}~\bibnamefont {Avouris}},
  \bibinfo {author} {\bibfnamefont {A.~C.}\ \bibnamefont {Ferrari}}, \bibinfo
  {author} {\bibfnamefont {M.~S.}\ \bibnamefont {Vitiello}}, \ and\ \bibinfo
  {author} {\bibfnamefont {M.}~\bibnamefont {Polini}},\ }\bibfield  {title}
  {\enquote {\bibinfo {title} {Photodetectors based on graphene, other
  two-dimensional materials and hybrid systems},}\ }\href {\doibase
  10.1038/nnano.2014.215} {\bibfield  {journal} {\bibinfo  {journal} {Nature
  Nanotechnology}\ }\textbf {\bibinfo {volume} {9}},\ \bibinfo {pages} {780 EP
  --} (\bibinfo {year} {2014})}\BibitemShut {NoStop}%
\bibitem [{\citenamefont {Saran}\ and\ \citenamefont
  {Curry}(2016)}]{Saran2016}%
  \BibitemOpen
  \bibfield  {author} {\bibinfo {author} {\bibfnamefont {R.}~\bibnamefont
  {Saran}}\ and\ \bibinfo {author} {\bibfnamefont {R.~J.}\ \bibnamefont
  {Curry}},\ }\bibfield  {title} {\enquote {\bibinfo {title} {Lead sulphide
  nanocrystal photodetector technologies},}\ }\href {\doibase
  10.1038/nphoton.2015.280} {\bibfield  {journal} {\bibinfo  {journal} {Nature
  Photonics}\ }\textbf {\bibinfo {volume} {10}},\ \bibinfo {pages} {81 EP --}
  (\bibinfo {year} {2016})}\BibitemShut {NoStop}%
\bibitem [{\citenamefont {Fang}\ and\ \citenamefont {Hu}(2017)}]{Fang2017}%
  \BibitemOpen
  \bibfield  {author} {\bibinfo {author} {\bibfnamefont {H.}~\bibnamefont
  {Fang}}\ and\ \bibinfo {author} {\bibfnamefont {W.}~\bibnamefont {Hu}},\
  }\bibfield  {title} {\enquote {\bibinfo {title} {Photogating in low
  dimensional photodetectors},}\ }\href {\doibase 10.1002/advs.201700323}
  {\bibfield  {journal} {\bibinfo  {journal} {Advanced Science}\ }\textbf
  {\bibinfo {volume} {4}},\ \bibinfo {pages} {1700323} (\bibinfo {year}
  {2017})}\BibitemShut {NoStop}%
\bibitem [{\citenamefont {Long}\ \emph {et~al.}(2018)\citenamefont {Long},
  \citenamefont {Wang}, \citenamefont {Fang},\ and\ \citenamefont
  {Hu}}]{Long2018}%
  \BibitemOpen
  \bibfield  {author} {\bibinfo {author} {\bibfnamefont {M.}~\bibnamefont
  {Long}}, \bibinfo {author} {\bibfnamefont {P.}~\bibnamefont {Wang}}, \bibinfo
  {author} {\bibfnamefont {H.}~\bibnamefont {Fang}}, \ and\ \bibinfo {author}
  {\bibfnamefont {W.}~\bibnamefont {Hu}},\ }\bibfield  {title} {\enquote
  {\bibinfo {title} {Progress, challenges, and opportunities for 2d material
  based photodetectors},}\ }\href {\doibase 10.1002/adfm.201803807} {\bibfield
  {journal} {\bibinfo  {journal} {Advanced Functional Materials}\ }\textbf
  {\bibinfo {volume} {29}},\ \bibinfo {pages} {1803807} (\bibinfo {year}
  {2018})}\BibitemShut {NoStop}%
\bibitem [{\citenamefont {Rana}\ \emph {et~al.}(2017)\citenamefont {Rana},
  \citenamefont {Lee}, \citenamefont {Shahid},\ and\ \citenamefont
  {Kim}}]{Rana2017}%
  \BibitemOpen
  \bibfield  {author} {\bibinfo {author} {\bibfnamefont {A.~U. H.~S.}\
  \bibnamefont {Rana}}, \bibinfo {author} {\bibfnamefont {J.~Y.}\ \bibnamefont
  {Lee}}, \bibinfo {author} {\bibfnamefont {A.}~\bibnamefont {Shahid}}, \ and\
  \bibinfo {author} {\bibfnamefont {H.-S.}\ \bibnamefont {Kim}},\ }\bibfield
  {title} {\enquote {\bibinfo {title} {Growth method-dependent and defect
  density-oriented structural, optical, conductive, and physical properties of
  solution-grown zno nanostructures},}\ }\href {\doibase 10.3390/nano7090266}
  {\bibfield  {journal} {\bibinfo  {journal} {Nanomaterials}\ }\textbf
  {\bibinfo {volume} {7}} (\bibinfo {year} {2017}),\
  10.3390/nano7090266}\BibitemShut {NoStop}%
\bibitem [{\citenamefont {Lou}\ \emph {et~al.}(2013)\citenamefont {Lou},
  \citenamefont {Yuan}, \citenamefont {Zhao}, \citenamefont {Wang},\ and\
  \citenamefont {Shi}}]{Lou2013}%
  \BibitemOpen
  \bibfield  {author} {\bibinfo {author} {\bibfnamefont {Y.-Y.}\ \bibnamefont
  {Lou}}, \bibinfo {author} {\bibfnamefont {S.}~\bibnamefont {Yuan}}, \bibinfo
  {author} {\bibfnamefont {Y.}~\bibnamefont {Zhao}}, \bibinfo {author}
  {\bibfnamefont {Z.-Y.}\ \bibnamefont {Wang}}, \ and\ \bibinfo {author}
  {\bibfnamefont {L.-Y.}\ \bibnamefont {Shi}},\ }\bibfield  {title} {\enquote
  {\bibinfo {title} {Influence of defect density on the zno nanostructures of
  dye-sensitized solar cells},}\ }\href
  {https://doi.org/10.1007/s40436-013-0046-x} {\bibfield  {journal} {\bibinfo
  {journal} {Advances in Manufacturing}\ }\textbf {\bibinfo {volume} {1}},\
  \bibinfo {pages} {340--345} (\bibinfo {year} {2013})}\BibitemShut {NoStop}%
\bibitem [{\citenamefont {Norris}\ and\ \citenamefont
  {Bawendi}(1996)}]{Norris1996}%
  \BibitemOpen
  \bibfield  {author} {\bibinfo {author} {\bibfnamefont {D.~J.}\ \bibnamefont
  {Norris}}\ and\ \bibinfo {author} {\bibfnamefont {M.~G.}\ \bibnamefont
  {Bawendi}},\ }\bibfield  {title} {\enquote {\bibinfo {title} {Measurement and
  assignment of the size-dependent optical spectrum in cdse quantum dots},}\
  }\href {\doibase 10.1103/PhysRevB.53.16338} {\bibfield  {journal} {\bibinfo
  {journal} {Phys. Rev. B}\ }\textbf {\bibinfo {volume} {53}},\ \bibinfo
  {pages} {16338--16346} (\bibinfo {year} {1996})}\BibitemShut {NoStop}%
\bibitem [{\citenamefont {Wu}\ \emph {et~al.}(1987)\citenamefont {Wu},
  \citenamefont {Schulman}, \citenamefont {Hsu},\ and\ \citenamefont
  {Efron}}]{Wu1987}%
  \BibitemOpen
  \bibfield  {author} {\bibinfo {author} {\bibfnamefont {W.}~\bibnamefont
  {Wu}}, \bibinfo {author} {\bibfnamefont {J.~N.}\ \bibnamefont {Schulman}},
  \bibinfo {author} {\bibfnamefont {T.~Y.}\ \bibnamefont {Hsu}}, \ and\
  \bibinfo {author} {\bibfnamefont {U.}~\bibnamefont {Efron}},\ }\bibfield
  {title} {\enquote {\bibinfo {title} {Effect of size nonuniformity on the
  absorption spectrum of a semiconductor quantum dot system},}\ }\href
  {\doibase 10.1063/1.98896} {\bibfield  {journal} {\bibinfo  {journal}
  {Applied Physics Letters}\ }\textbf {\bibinfo {volume} {51}},\ \bibinfo
  {pages} {710--712} (\bibinfo {year} {1987})}\BibitemShut {NoStop}%
\bibitem [{\citenamefont {Kongkanand}\ \emph {et~al.}(2008)\citenamefont
  {Kongkanand}, \citenamefont {Tvrdy}, \citenamefont {Takechi}, \citenamefont
  {Kuno},\ and\ \citenamefont {Kamat}}]{Kongkanand2008}%
  \BibitemOpen
  \bibfield  {author} {\bibinfo {author} {\bibfnamefont {A.}~\bibnamefont
  {Kongkanand}}, \bibinfo {author} {\bibfnamefont {K.}~\bibnamefont {Tvrdy}},
  \bibinfo {author} {\bibfnamefont {K.}~\bibnamefont {Takechi}}, \bibinfo
  {author} {\bibfnamefont {M.}~\bibnamefont {Kuno}}, \ and\ \bibinfo {author}
  {\bibfnamefont {P.~V.}\ \bibnamefont {Kamat}},\ }\bibfield  {title} {\enquote
  {\bibinfo {title} {Quantum dot solar cells. tuning photoresponse through size
  and shape control of cdse-tio2 architecture},}\ }\href {\doibase
  10.1021/ja0782706} {\bibfield  {journal} {\bibinfo  {journal} {Journal of the
  American Chemical Society}\ }\textbf {\bibinfo {volume} {130}},\ \bibinfo
  {pages} {4007--4015} (\bibinfo {year} {2008})},\ \bibinfo {note} {pMID:
  18311974}\BibitemShut {NoStop}%
\bibitem [{\citenamefont {Rahneshin}\ \emph {et~al.}(2016)\citenamefont
  {Rahneshin}, \citenamefont {Khosravi}, \citenamefont {Ziolkowska},
  \citenamefont {Jasinski},\ and\ \citenamefont
  {Panchapakesan}}]{Rahneshin2016}%
  \BibitemOpen
  \bibfield  {author} {\bibinfo {author} {\bibfnamefont {V.}~\bibnamefont
  {Rahneshin}}, \bibinfo {author} {\bibfnamefont {F.}~\bibnamefont {Khosravi}},
  \bibinfo {author} {\bibfnamefont {D.~A.}\ \bibnamefont {Ziolkowska}},
  \bibinfo {author} {\bibfnamefont {J.~B.}\ \bibnamefont {Jasinski}}, \ and\
  \bibinfo {author} {\bibfnamefont {B.}~\bibnamefont {Panchapakesan}},\
  }\bibfield  {title} {\enquote {\bibinfo {title} {Chromatic mechanical
  response in 2-d layered transition metal dichalcogenide (tmds) based
  nanocomposites},}\ }\href {https://doi.org/10.1038/srep34831} {\bibfield
  {journal} {\bibinfo  {journal} {Scientific Reports}\ }\textbf {\bibinfo
  {volume} {6}},\ \bibinfo {pages} {34831 EP --} (\bibinfo {year} {2016})},\
  \bibinfo {note} {article}\BibitemShut {NoStop}%
\bibitem [{\citenamefont {Lee}\ \emph {et~al.}(2017)\citenamefont {Lee},
  \citenamefont {Huang}, \citenamefont {Sumpter},\ and\ \citenamefont
  {Yoon}}]{Lee2017}%
  \BibitemOpen
  \bibfield  {author} {\bibinfo {author} {\bibfnamefont {J.}~\bibnamefont
  {Lee}}, \bibinfo {author} {\bibfnamefont {J.}~\bibnamefont {Huang}}, \bibinfo
  {author} {\bibfnamefont {B.~G.}\ \bibnamefont {Sumpter}}, \ and\ \bibinfo
  {author} {\bibfnamefont {M.}~\bibnamefont {Yoon}},\ }\bibfield  {title}
  {\enquote {\bibinfo {title} {Strain-engineered optoelectronic properties of
  2d transition metal dichalcogenide lateral heterostructures},}\ }\href
  {\doibase 10.1088/2053-1583/aa5542} {\bibfield  {journal} {\bibinfo
  {journal} {2D Materials}\ }\textbf {\bibinfo {volume} {4}},\ \bibinfo {pages}
  {021016} (\bibinfo {year} {2017})}\BibitemShut {NoStop}%
\bibitem [{\citenamefont {Miao}\ \emph {et~al.}(2019)\citenamefont {Miao},
  \citenamefont {Shu}, \citenamefont {Hu}, \citenamefont {Tang}, \citenamefont
  {Hao}, \citenamefont {You}, \citenamefont {Zheng}, \citenamefont {Cheng},
  \citenamefont {Duan},\ and\ \citenamefont {Jiang}}]{Miao2019}%
  \BibitemOpen
  \bibfield  {author} {\bibinfo {author} {\bibfnamefont {R.}~\bibnamefont
  {Miao}}, \bibinfo {author} {\bibfnamefont {Z.}~\bibnamefont {Shu}}, \bibinfo
  {author} {\bibfnamefont {Y.}~\bibnamefont {Hu}}, \bibinfo {author}
  {\bibfnamefont {Y.}~\bibnamefont {Tang}}, \bibinfo {author} {\bibfnamefont
  {H.}~\bibnamefont {Hao}}, \bibinfo {author} {\bibfnamefont {J.}~\bibnamefont
  {You}}, \bibinfo {author} {\bibfnamefont {X.}~\bibnamefont {Zheng}}, \bibinfo
  {author} {\bibfnamefont {X.}~\bibnamefont {Cheng}}, \bibinfo {author}
  {\bibfnamefont {H.}~\bibnamefont {Duan}}, \ and\ \bibinfo {author}
  {\bibfnamefont {T.}~\bibnamefont {Jiang}},\ }\bibfield  {title} {\enquote
  {\bibinfo {title} {Ultrafast nonlinear absorption enhancement of monolayer
  mos2 with plasmonic au nanoantennas},}\ }\href {\doibase
  10.1364/OL.44.003198} {\bibfield  {journal} {\bibinfo  {journal} {Opt.
  Lett.}\ }\textbf {\bibinfo {volume} {44}},\ \bibinfo {pages} {3198--3201}
  (\bibinfo {year} {2019})}\BibitemShut {NoStop}%
\bibitem [{\citenamefont {Lin}\ \emph {et~al.}(2019)\citenamefont {Lin},
  \citenamefont {Ye}, \citenamefont {Chen}, \citenamefont {Zhou}, \citenamefont
  {Yi}, \citenamefont {Niu}, \citenamefont {Yi}, \citenamefont {Hua},
  \citenamefont {Hua},\ and\ \citenamefont {Xiao}}]{Lin2019}%
  \BibitemOpen
  \bibfield  {author} {\bibinfo {author} {\bibfnamefont {H.}~\bibnamefont
  {Lin}}, \bibinfo {author} {\bibfnamefont {X.}~\bibnamefont {Ye}}, \bibinfo
  {author} {\bibfnamefont {X.}~\bibnamefont {Chen}}, \bibinfo {author}
  {\bibfnamefont {Z.}~\bibnamefont {Zhou}}, \bibinfo {author} {\bibfnamefont
  {Z.}~\bibnamefont {Yi}}, \bibinfo {author} {\bibfnamefont {G.}~\bibnamefont
  {Niu}}, \bibinfo {author} {\bibfnamefont {Y.}~\bibnamefont {Yi}}, \bibinfo
  {author} {\bibfnamefont {Y.}~\bibnamefont {Hua}}, \bibinfo {author}
  {\bibfnamefont {J.}~\bibnamefont {Hua}}, \ and\ \bibinfo {author}
  {\bibfnamefont {S.}~\bibnamefont {Xiao}},\ }\bibfield  {title} {\enquote
  {\bibinfo {title} {Plasmonic absorption enhancement in graphene circular and
  elliptical disk arrays},}\ }\href {\doibase 10.1088/2053-1591/aafc3e}
  {\bibfield  {journal} {\bibinfo  {journal} {Materials Research Express}\
  }\textbf {\bibinfo {volume} {6}},\ \bibinfo {pages} {045807} (\bibinfo {year}
  {2019})}\BibitemShut {NoStop}%
\bibitem [{\citenamefont {Zhou}\ \emph {et~al.}(2018)\citenamefont {Zhou},
  \citenamefont {Tan}, \citenamefont {Sheng}, \citenamefont {Fan},
  \citenamefont {Xu},\ and\ \citenamefont {Warner}}]{Zhou2018}%
  \BibitemOpen
  \bibfield  {author} {\bibinfo {author} {\bibfnamefont {Y.}~\bibnamefont
  {Zhou}}, \bibinfo {author} {\bibfnamefont {H.}~\bibnamefont {Tan}}, \bibinfo
  {author} {\bibfnamefont {Y.}~\bibnamefont {Sheng}}, \bibinfo {author}
  {\bibfnamefont {Y.}~\bibnamefont {Fan}}, \bibinfo {author} {\bibfnamefont
  {W.}~\bibnamefont {Xu}}, \ and\ \bibinfo {author} {\bibfnamefont {J.~H.}\
  \bibnamefont {Warner}},\ }\bibfield  {title} {\enquote {\bibinfo {title}
  {Utilizing interlayer excitons in bilayer ws2 for increased photovoltaic
  response in ultrathin graphene vertical cross-bar photodetecting tunneling
  transistors},}\ }\href {\doibase 10.1021/acsnano.8b01263} {\bibfield
  {journal} {\bibinfo  {journal} {ACS Nano}\ }\textbf {\bibinfo {volume}
  {12}},\ \bibinfo {pages} {4669--4677} (\bibinfo {year} {2018})},\ \bibinfo
  {note} {pMID: 29671322}\BibitemShut {NoStop}%
\bibitem [{\citenamefont {Long}\ \emph {et~al.}(2016)\citenamefont {Long},
  \citenamefont {Liu}, \citenamefont {Wang}, \citenamefont {Gao}, \citenamefont
  {Xia}, \citenamefont {Luo}, \citenamefont {Wang}, \citenamefont {Zeng},
  \citenamefont {Fu}, \citenamefont {Xu}, \citenamefont {Zhou}, \citenamefont
  {Lv}, \citenamefont {Yao}, \citenamefont {Lu}, \citenamefont {Chen},
  \citenamefont {Ni}, \citenamefont {You}, \citenamefont {Zhang}, \citenamefont
  {Qin}, \citenamefont {Shi}, \citenamefont {Hu}, \citenamefont {Xing},\ and\
  \citenamefont {Miao}}]{Long2016}%
  \BibitemOpen
  \bibfield  {author} {\bibinfo {author} {\bibfnamefont {M.}~\bibnamefont
  {Long}}, \bibinfo {author} {\bibfnamefont {E.}~\bibnamefont {Liu}}, \bibinfo
  {author} {\bibfnamefont {P.}~\bibnamefont {Wang}}, \bibinfo {author}
  {\bibfnamefont {A.}~\bibnamefont {Gao}}, \bibinfo {author} {\bibfnamefont
  {H.}~\bibnamefont {Xia}}, \bibinfo {author} {\bibfnamefont {W.}~\bibnamefont
  {Luo}}, \bibinfo {author} {\bibfnamefont {B.}~\bibnamefont {Wang}}, \bibinfo
  {author} {\bibfnamefont {J.}~\bibnamefont {Zeng}}, \bibinfo {author}
  {\bibfnamefont {Y.}~\bibnamefont {Fu}}, \bibinfo {author} {\bibfnamefont
  {K.}~\bibnamefont {Xu}}, \bibinfo {author} {\bibfnamefont {W.}~\bibnamefont
  {Zhou}}, \bibinfo {author} {\bibfnamefont {Y.}~\bibnamefont {Lv}}, \bibinfo
  {author} {\bibfnamefont {S.}~\bibnamefont {Yao}}, \bibinfo {author}
  {\bibfnamefont {M.}~\bibnamefont {Lu}}, \bibinfo {author} {\bibfnamefont
  {Y.}~\bibnamefont {Chen}}, \bibinfo {author} {\bibfnamefont {Z.}~\bibnamefont
  {Ni}}, \bibinfo {author} {\bibfnamefont {Y.}~\bibnamefont {You}}, \bibinfo
  {author} {\bibfnamefont {X.}~\bibnamefont {Zhang}}, \bibinfo {author}
  {\bibfnamefont {S.}~\bibnamefont {Qin}}, \bibinfo {author} {\bibfnamefont
  {Y.}~\bibnamefont {Shi}}, \bibinfo {author} {\bibfnamefont {W.}~\bibnamefont
  {Hu}}, \bibinfo {author} {\bibfnamefont {D.}~\bibnamefont {Xing}}, \ and\
  \bibinfo {author} {\bibfnamefont {F.}~\bibnamefont {Miao}},\ }\bibfield
  {title} {\enquote {\bibinfo {title} {Broadband photovoltaic detectors based
  on an atomically thin heterostructure},}\ }\href {\doibase
  10.1021/acs.nanolett.5b04538} {\bibfield  {journal} {\bibinfo  {journal}
  {Nano Letters}\ }\textbf {\bibinfo {volume} {16}},\ \bibinfo {pages}
  {2254--2259} (\bibinfo {year} {2016})},\ \bibinfo {note} {pMID:
  26886761}\BibitemShut {NoStop}%
\bibitem [{\citenamefont {Massicotte}\ \emph {et~al.}(2015)\citenamefont
  {Massicotte}, \citenamefont {Schmidt}, \citenamefont {Vialla}, \citenamefont
  {Sch{\"a}dler}, \citenamefont {Reserbat-Plantey}, \citenamefont {Watanabe},
  \citenamefont {Taniguchi}, \citenamefont {Tielrooij},\ and\ \citenamefont
  {Koppens}}]{Koppens2015}%
  \BibitemOpen
  \bibfield  {author} {\bibinfo {author} {\bibfnamefont {M.}~\bibnamefont
  {Massicotte}}, \bibinfo {author} {\bibfnamefont {P.}~\bibnamefont {Schmidt}},
  \bibinfo {author} {\bibfnamefont {F.}~\bibnamefont {Vialla}}, \bibinfo
  {author} {\bibfnamefont {K.~G.}\ \bibnamefont {Sch{\"a}dler}}, \bibinfo
  {author} {\bibfnamefont {A.}~\bibnamefont {Reserbat-Plantey}}, \bibinfo
  {author} {\bibfnamefont {K.}~\bibnamefont {Watanabe}}, \bibinfo {author}
  {\bibfnamefont {T.}~\bibnamefont {Taniguchi}}, \bibinfo {author}
  {\bibfnamefont {K.~J.}\ \bibnamefont {Tielrooij}}, \ and\ \bibinfo {author}
  {\bibfnamefont {F.~H.~L.}\ \bibnamefont {Koppens}},\ }\bibfield  {title}
  {\enquote {\bibinfo {title} {Picosecond photoresponse in van der waals
  heterostructures},}\ }\href {https://doi.org/10.1038/nnano.2015.227}
  {\bibfield  {journal} {\bibinfo  {journal} {Nature Nanotechnology}\ }\textbf
  {\bibinfo {volume} {11}},\ \bibinfo {pages} {42 EP --} (\bibinfo {year}
  {2015})}\BibitemShut {NoStop}%
\bibitem [{\citenamefont {Zhang}\ \emph {et~al.}(2017)\citenamefont {Zhang},
  \citenamefont {Fang}, \citenamefont {Wang}, \citenamefont {Wan},
  \citenamefont {Song}, \citenamefont {Zhai}, \citenamefont {Li}, \citenamefont
  {Ran}, \citenamefont {Ye},\ and\ \citenamefont {Dai}}]{Zhang2017}%
  \BibitemOpen
  \bibfield  {author} {\bibinfo {author} {\bibfnamefont {K.}~\bibnamefont
  {Zhang}}, \bibinfo {author} {\bibfnamefont {X.}~\bibnamefont {Fang}},
  \bibinfo {author} {\bibfnamefont {Y.}~\bibnamefont {Wang}}, \bibinfo {author}
  {\bibfnamefont {Y.}~\bibnamefont {Wan}}, \bibinfo {author} {\bibfnamefont
  {Q.}~\bibnamefont {Song}}, \bibinfo {author} {\bibfnamefont {W.}~\bibnamefont
  {Zhai}}, \bibinfo {author} {\bibfnamefont {Y.}~\bibnamefont {Li}}, \bibinfo
  {author} {\bibfnamefont {G.}~\bibnamefont {Ran}}, \bibinfo {author}
  {\bibfnamefont {Y.}~\bibnamefont {Ye}}, \ and\ \bibinfo {author}
  {\bibfnamefont {L.}~\bibnamefont {Dai}},\ }\bibfield  {title} {\enquote
  {\bibinfo {title} {Ultrasensitive near-infrared photodetectors based on a
  graphene–mote2–graphene vertical van der waals heterostructure},}\ }\href
  {\doibase 10.1021/acsami.6b14483} {\bibfield  {journal} {\bibinfo  {journal}
  {ACS Applied Materials \& Interfaces}\ }\textbf {\bibinfo {volume} {9}},\
  \bibinfo {pages} {5392--5398} (\bibinfo {year} {2017})},\ \bibinfo {note}
  {pMID: 28111947}\BibitemShut {NoStop}%
\bibitem [{\citenamefont {Xia}\ \emph {et~al.}(2009)\citenamefont {Xia},
  \citenamefont {Mueller}, \citenamefont {Lin}, \citenamefont {Valdes-Garcia},\
  and\ \citenamefont {Avouris}}]{Xia2009}%
  \BibitemOpen
  \bibfield  {author} {\bibinfo {author} {\bibfnamefont {F.}~\bibnamefont
  {Xia}}, \bibinfo {author} {\bibfnamefont {T.}~\bibnamefont {Mueller}},
  \bibinfo {author} {\bibfnamefont {Y.-m.}\ \bibnamefont {Lin}}, \bibinfo
  {author} {\bibfnamefont {A.}~\bibnamefont {Valdes-Garcia}}, \ and\ \bibinfo
  {author} {\bibfnamefont {P.}~\bibnamefont {Avouris}},\ }\bibfield  {title}
  {\enquote {\bibinfo {title} {Ultrafast graphene photodetector},}\ }\href
  {\doibase 10.1038/nnano.2009.292} {\bibfield  {journal} {\bibinfo  {journal}
  {Nature Nanotechnology}\ }\textbf {\bibinfo {volume} {4}},\ \bibinfo {pages}
  {839--843} (\bibinfo {year} {2009})}\BibitemShut {NoStop}%
\bibitem [{\citenamefont {Kufer}\ and\ \citenamefont
  {Konstantatos}(2015)}]{Konstantatos2015}%
  \BibitemOpen
  \bibfield  {author} {\bibinfo {author} {\bibfnamefont {D.}~\bibnamefont
  {Kufer}}\ and\ \bibinfo {author} {\bibfnamefont {G.}~\bibnamefont
  {Konstantatos}},\ }\bibfield  {title} {\enquote {\bibinfo {title} {Highly
  sensitive, encapsulated mos2 photodetector with gate controllable gain and
  speed},}\ }\href {\doibase 10.1021/acs.nanolett.5b02559} {\bibfield
  {journal} {\bibinfo  {journal} {Nano Letters}\ }\textbf {\bibinfo {volume}
  {15}},\ \bibinfo {pages} {7307--7313} (\bibinfo {year} {2015})},\ \bibinfo
  {note} {pMID: 26501356}\BibitemShut {NoStop}%
\bibitem [{\citenamefont {Lopez-Sanchez}\ \emph
  {et~al.}(2013{\natexlab{a}})\citenamefont {Lopez-Sanchez}, \citenamefont
  {Lembke}, \citenamefont {Kayci}, \citenamefont {Radenovic},\ and\
  \citenamefont {Kis}}]{Kis2013}%
  \BibitemOpen
  \bibfield  {author} {\bibinfo {author} {\bibfnamefont {O.}~\bibnamefont
  {Lopez-Sanchez}}, \bibinfo {author} {\bibfnamefont {D.}~\bibnamefont
  {Lembke}}, \bibinfo {author} {\bibfnamefont {M.}~\bibnamefont {Kayci}},
  \bibinfo {author} {\bibfnamefont {A.}~\bibnamefont {Radenovic}}, \ and\
  \bibinfo {author} {\bibfnamefont {A.}~\bibnamefont {Kis}},\ }\bibfield
  {title} {\enquote {\bibinfo {title} {Ultrasensitive photodetectors based on
  monolayer mos2},}\ }\href {https://doi.org/10.1038/nnano.2013.100} {\bibfield
   {journal} {\bibinfo  {journal} {Nature Nanotechnology}\ }\textbf {\bibinfo
  {volume} {8}},\ \bibinfo {pages} {497 EP --} (\bibinfo {year}
  {2013}{\natexlab{a}})}\BibitemShut {NoStop}%
\bibitem [{\citenamefont {Li}\ \emph {et~al.}(2019)\citenamefont {Li},
  \citenamefont {Lan}, \citenamefont {Manikandan}, \citenamefont {Yip},
  \citenamefont {Zhou}, \citenamefont {Liang}, \citenamefont {Shu},
  \citenamefont {Chueh}, \citenamefont {Han},\ and\ \citenamefont
  {Ho}}]{Li2019}%
  \BibitemOpen
  \bibfield  {author} {\bibinfo {author} {\bibfnamefont {D.}~\bibnamefont
  {Li}}, \bibinfo {author} {\bibfnamefont {C.}~\bibnamefont {Lan}}, \bibinfo
  {author} {\bibfnamefont {A.}~\bibnamefont {Manikandan}}, \bibinfo {author}
  {\bibfnamefont {S.}~\bibnamefont {Yip}}, \bibinfo {author} {\bibfnamefont
  {Z.}~\bibnamefont {Zhou}}, \bibinfo {author} {\bibfnamefont {X.}~\bibnamefont
  {Liang}}, \bibinfo {author} {\bibfnamefont {L.}~\bibnamefont {Shu}}, \bibinfo
  {author} {\bibfnamefont {Y.-L.}\ \bibnamefont {Chueh}}, \bibinfo {author}
  {\bibfnamefont {N.}~\bibnamefont {Han}}, \ and\ \bibinfo {author}
  {\bibfnamefont {J.~C.}\ \bibnamefont {Ho}},\ }\bibfield  {title} {\enquote
  {\bibinfo {title} {Ultra-fast photodetectors based on high-mobility indium
  gallium antimonide nanowires},}\ }\href {\doibase 10.1038/s41467-019-09606-y}
  {\bibfield  {journal} {\bibinfo  {journal} {Nature Communications}\ }\textbf
  {\bibinfo {volume} {10}},\ \bibinfo {pages} {1664} (\bibinfo {year}
  {2019})}\BibitemShut {NoStop}%
\bibitem [{\citenamefont {Luo}\ \emph {et~al.}(2018{\natexlab{a}})\citenamefont
  {Luo}, \citenamefont {Zhu}, \citenamefont {tan}, \citenamefont {Sun},
  \citenamefont {Luo}, \citenamefont {Peng}, \citenamefont {Zhu}, \citenamefont
  {Zhang},\ and\ \citenamefont {Qin}}]{Luo20182}%
  \BibitemOpen
  \bibfield  {author} {\bibinfo {author} {\bibfnamefont {F.}~\bibnamefont
  {Luo}}, \bibinfo {author} {\bibfnamefont {M.}~\bibnamefont {Zhu}}, \bibinfo
  {author} {\bibfnamefont {Y.}~\bibnamefont {tan}}, \bibinfo {author}
  {\bibfnamefont {H.}~\bibnamefont {Sun}}, \bibinfo {author} {\bibfnamefont
  {W.}~\bibnamefont {Luo}}, \bibinfo {author} {\bibfnamefont {G.}~\bibnamefont
  {Peng}}, \bibinfo {author} {\bibfnamefont {Z.}~\bibnamefont {Zhu}}, \bibinfo
  {author} {\bibfnamefont {X.-A.}\ \bibnamefont {Zhang}}, \ and\ \bibinfo
  {author} {\bibfnamefont {S.}~\bibnamefont {Qin}},\ }\bibfield  {title}
  {\enquote {\bibinfo {title} {High responsivity graphene photodetectors from
  visible to near-infrared by photogating effect},}\ }\href {\doibase
  10.1063/1.5054760} {\bibfield  {journal} {\bibinfo  {journal} {AIP Advances}\
  }\textbf {\bibinfo {volume} {8}},\ \bibinfo {pages} {115106} (\bibinfo {year}
  {2018}{\natexlab{a}})}\BibitemShut {NoStop}%
\bibitem [{\citenamefont {Zhang}\ \emph {et~al.}(2019)\citenamefont {Zhang},
  \citenamefont {Peng}, \citenamefont {Yu}, \citenamefont {Fan}, \citenamefont
  {Zhai},\ and\ \citenamefont {Wang}}]{Zhang2019}%
  \BibitemOpen
  \bibfield  {author} {\bibinfo {author} {\bibfnamefont {K.}~\bibnamefont
  {Zhang}}, \bibinfo {author} {\bibfnamefont {M.}~\bibnamefont {Peng}},
  \bibinfo {author} {\bibfnamefont {A.}~\bibnamefont {Yu}}, \bibinfo {author}
  {\bibfnamefont {Y.}~\bibnamefont {Fan}}, \bibinfo {author} {\bibfnamefont
  {J.}~\bibnamefont {Zhai}}, \ and\ \bibinfo {author} {\bibfnamefont {Z.~L.}\
  \bibnamefont {Wang}},\ }\bibfield  {title} {\enquote {\bibinfo {title} {A
  substrate-enhanced mos2 photodetector through a dual-photogating effect},}\
  }\href {\doibase 10.1039/C8MH01429A} {\bibfield  {journal} {\bibinfo
  {journal} {Mater. Horiz.}\ }\textbf {\bibinfo {volume} {6}},\ \bibinfo
  {pages} {826--833} (\bibinfo {year} {2019})}\BibitemShut {NoStop}%
\bibitem [{\citenamefont {Liu}\ \emph {et~al.}(2019{\natexlab{a}})\citenamefont
  {Liu}, \citenamefont {Zhao}, \citenamefont {Chen}, \citenamefont {Zhang},
  \citenamefont {Li}, \citenamefont {Yan},\ and\ \citenamefont
  {Zhang}}]{Liu2019}%
  \BibitemOpen
  \bibfield  {author} {\bibinfo {author} {\bibfnamefont {B.}~\bibnamefont
  {Liu}}, \bibinfo {author} {\bibfnamefont {C.}~\bibnamefont {Zhao}}, \bibinfo
  {author} {\bibfnamefont {X.}~\bibnamefont {Chen}}, \bibinfo {author}
  {\bibfnamefont {L.}~\bibnamefont {Zhang}}, \bibinfo {author} {\bibfnamefont
  {Y.}~\bibnamefont {Li}}, \bibinfo {author} {\bibfnamefont {H.}~\bibnamefont
  {Yan}}, \ and\ \bibinfo {author} {\bibfnamefont {Y.}~\bibnamefont {Zhang}},\
  }\bibfield  {title} {\enquote {\bibinfo {title} {Self-powered and fast
  photodetector based on graphene/mose2/au heterojunction},}\ }\href {\doibase
  https://doi.org/10.1016/j.spmi.2019.04.021} {\bibfield  {journal} {\bibinfo
  {journal} {Superlattices and Microstructures}\ }\textbf {\bibinfo {volume}
  {130}},\ \bibinfo {pages} {87 -- 92} (\bibinfo {year}
  {2019}{\natexlab{a}})}\BibitemShut {NoStop}%
\bibitem [{\citenamefont {Liu}\ \emph {et~al.}(2018)\citenamefont {Liu},
  \citenamefont {Li}, \citenamefont {Xu},\ and\ \citenamefont {Qi}}]{Liu2018}%
  \BibitemOpen
  \bibfield  {author} {\bibinfo {author} {\bibfnamefont {X.}~\bibnamefont
  {Liu}}, \bibinfo {author} {\bibfnamefont {F.}~\bibnamefont {Li}}, \bibinfo
  {author} {\bibfnamefont {M.}~\bibnamefont {Xu}}, \ and\ \bibinfo {author}
  {\bibfnamefont {J.}~\bibnamefont {Qi}},\ }\bibfield  {title} {\enquote
  {\bibinfo {title} {Self-powered{,} high response and fast response speed
  metal–insulator–semiconductor structured photodetector based on 2d
  mos2},}\ }\href {\doibase 10.1039/C8RA05511D} {\bibfield  {journal} {\bibinfo
   {journal} {RSC Adv.}\ }\textbf {\bibinfo {volume} {8}},\ \bibinfo {pages}
  {28041--28047} (\bibinfo {year} {2018})}\BibitemShut {NoStop}%
\bibitem [{\citenamefont {Viti}\ \emph {et~al.}(2019)\citenamefont {Viti},
  \citenamefont {Politano}, \citenamefont {Zhang},\ and\ \citenamefont
  {Vitiello}}]{Viti2019}%
  \BibitemOpen
  \bibfield  {author} {\bibinfo {author} {\bibfnamefont {L.}~\bibnamefont
  {Viti}}, \bibinfo {author} {\bibfnamefont {A.}~\bibnamefont {Politano}},
  \bibinfo {author} {\bibfnamefont {K.}~\bibnamefont {Zhang}}, \ and\ \bibinfo
  {author} {\bibfnamefont {M.~S.}\ \bibnamefont {Vitiello}},\ }\bibfield
  {title} {\enquote {\bibinfo {title} {Thermoelectric terahertz photodetectors
  based on selenium-doped black phosphorus flakes},}\ }\href {\doibase
  10.1039/C8NR09060B} {\bibfield  {journal} {\bibinfo  {journal} {Nanoscale}\
  }\textbf {\bibinfo {volume} {11}},\ \bibinfo {pages} {1995--2002} (\bibinfo
  {year} {2019})}\BibitemShut {NoStop}%
\bibitem [{\citenamefont {Ryzhii}\ \emph {et~al.}(2019)\citenamefont {Ryzhii},
  \citenamefont {Ryzhii}, \citenamefont {Ponomarev}, \citenamefont {Leiman},
  \citenamefont {Mitin}, \citenamefont {Shur},\ and\ \citenamefont
  {Otsuji}}]{Ryzhii2019}%
  \BibitemOpen
  \bibfield  {author} {\bibinfo {author} {\bibfnamefont {V.}~\bibnamefont
  {Ryzhii}}, \bibinfo {author} {\bibfnamefont {M.}~\bibnamefont {Ryzhii}},
  \bibinfo {author} {\bibfnamefont {D.~S.}\ \bibnamefont {Ponomarev}}, \bibinfo
  {author} {\bibfnamefont {V.~G.}\ \bibnamefont {Leiman}}, \bibinfo {author}
  {\bibfnamefont {V.}~\bibnamefont {Mitin}}, \bibinfo {author} {\bibfnamefont
  {M.~S.}\ \bibnamefont {Shur}}, \ and\ \bibinfo {author} {\bibfnamefont
  {T.}~\bibnamefont {Otsuji}},\ }\bibfield  {title} {\enquote {\bibinfo {title}
  {Negative photoconductivity and hot-carrier bolometric detection of terahertz
  radiation in graphene-phosphorene hybrid structures},}\ }\href {\doibase
  10.1063/1.5054142} {\bibfield  {journal} {\bibinfo  {journal} {Journal of
  Applied Physics}\ }\textbf {\bibinfo {volume} {125}},\ \bibinfo {pages}
  {151608} (\bibinfo {year} {2019})}\BibitemShut {NoStop}%
\bibitem [{\citenamefont {Liu}\ \emph {et~al.}(2019{\natexlab{b}})\citenamefont
  {Liu}, \citenamefont {Hu}, \citenamefont {Yin}, \citenamefont {Wang},
  \citenamefont {Wang}, \citenamefont {Wen}, \citenamefont {Dong},
  \citenamefont {Zhu}, \citenamefont {Wei}, \citenamefont {Ma},\ and\
  \citenamefont {Sun}}]{Liu20192}%
  \BibitemOpen
  \bibfield  {author} {\bibinfo {author} {\bibfnamefont {Y.}~\bibnamefont
  {Liu}}, \bibinfo {author} {\bibfnamefont {Q.}~\bibnamefont {Hu}}, \bibinfo
  {author} {\bibfnamefont {J.}~\bibnamefont {Yin}}, \bibinfo {author}
  {\bibfnamefont {P.}~\bibnamefont {Wang}}, \bibinfo {author} {\bibfnamefont
  {Y.}~\bibnamefont {Wang}}, \bibinfo {author} {\bibfnamefont {J.}~\bibnamefont
  {Wen}}, \bibinfo {author} {\bibfnamefont {Z.}~\bibnamefont {Dong}}, \bibinfo
  {author} {\bibfnamefont {J.-L.}\ \bibnamefont {Zhu}}, \bibinfo {author}
  {\bibfnamefont {J.}~\bibnamefont {Wei}}, \bibinfo {author} {\bibfnamefont
  {W.}~\bibnamefont {Ma}}, \ and\ \bibinfo {author} {\bibfnamefont {J.-L.}\
  \bibnamefont {Sun}},\ }\bibfield  {title} {\enquote {\bibinfo {title}
  {Bolometric terahertz detection based on suspended carbon nanotube fibers},}\
  }\href {\doibase 10.7567/1882-0786/ab3bf1} {\bibfield  {journal} {\bibinfo
  {journal} {Applied Physics Express}\ }\textbf {\bibinfo {volume} {12}},\
  \bibinfo {pages} {096505} (\bibinfo {year} {2019}{\natexlab{b}})}\BibitemShut
  {NoStop}%
\bibitem [{\citenamefont {Sett}\ \emph {et~al.}(2018)\citenamefont {Sett},
  \citenamefont {Sengupta}, \citenamefont {Ganesh}, \citenamefont {Narayan},\
  and\ \citenamefont {Raychaudhuri}}]{Sett2018}%
  \BibitemOpen
  \bibfield  {author} {\bibinfo {author} {\bibfnamefont {S.}~\bibnamefont
  {Sett}}, \bibinfo {author} {\bibfnamefont {S.}~\bibnamefont {Sengupta}},
  \bibinfo {author} {\bibfnamefont {N.}~\bibnamefont {Ganesh}}, \bibinfo
  {author} {\bibfnamefont {K.~S.}\ \bibnamefont {Narayan}}, \ and\ \bibinfo
  {author} {\bibfnamefont {A.~K.}\ \bibnamefont {Raychaudhuri}},\ }\bibfield
  {title} {\enquote {\bibinfo {title} {Self-powered single semiconductor
  nanowire photodetector},}\ }\href {\doibase 10.1088/1361-6528/aada2d}
  {\bibfield  {journal} {\bibinfo  {journal} {Nanotechnology}\ }\textbf
  {\bibinfo {volume} {29}},\ \bibinfo {pages} {445202} (\bibinfo {year}
  {2018})}\BibitemShut {NoStop}%
\bibitem [{\citenamefont {Butanovs}\ \emph {et~al.}(2018)\citenamefont
  {Butanovs}, \citenamefont {Vlassov}, \citenamefont {Kuzmin}, \citenamefont
  {Piskunov}, \citenamefont {Butikova},\ and\ \citenamefont
  {Polyakov}}]{Butanovs2018}%
  \BibitemOpen
  \bibfield  {author} {\bibinfo {author} {\bibfnamefont {E.}~\bibnamefont
  {Butanovs}}, \bibinfo {author} {\bibfnamefont {S.}~\bibnamefont {Vlassov}},
  \bibinfo {author} {\bibfnamefont {A.}~\bibnamefont {Kuzmin}}, \bibinfo
  {author} {\bibfnamefont {S.}~\bibnamefont {Piskunov}}, \bibinfo {author}
  {\bibfnamefont {J.}~\bibnamefont {Butikova}}, \ and\ \bibinfo {author}
  {\bibfnamefont {B.}~\bibnamefont {Polyakov}},\ }\bibfield  {title} {\enquote
  {\bibinfo {title} {Fast-response single-nanowire photodetector based on
  zno/ws2 core/shell heterostructures},}\ }\href {\doibase
  10.1021/acsami.8b02241} {\bibfield  {journal} {\bibinfo  {journal} {ACS
  Applied Materials \& Interfaces}\ }\textbf {\bibinfo {volume} {10}},\
  \bibinfo {pages} {13869--13876} (\bibinfo {year} {2018})},\ \bibinfo {note}
  {pMID: 29619827}\BibitemShut {NoStop}%
\bibitem [{\citenamefont {Luo}\ \emph {et~al.}(2018{\natexlab{b}})\citenamefont
  {Luo}, \citenamefont {Weng}, \citenamefont {Long}, \citenamefont {Wang},
  \citenamefont {Gong}, \citenamefont {Fang}, \citenamefont {Luo},
  \citenamefont {Wang}, \citenamefont {Wang}, \citenamefont {Zheng},
  \citenamefont {Hu}, \citenamefont {Chen},\ and\ \citenamefont
  {Lu}}]{Luo2018}%
  \BibitemOpen
  \bibfield  {author} {\bibinfo {author} {\bibfnamefont {W.}~\bibnamefont
  {Luo}}, \bibinfo {author} {\bibfnamefont {Q.}~\bibnamefont {Weng}}, \bibinfo
  {author} {\bibfnamefont {M.}~\bibnamefont {Long}}, \bibinfo {author}
  {\bibfnamefont {P.}~\bibnamefont {Wang}}, \bibinfo {author} {\bibfnamefont
  {F.}~\bibnamefont {Gong}}, \bibinfo {author} {\bibfnamefont {H.}~\bibnamefont
  {Fang}}, \bibinfo {author} {\bibfnamefont {M.}~\bibnamefont {Luo}}, \bibinfo
  {author} {\bibfnamefont {W.}~\bibnamefont {Wang}}, \bibinfo {author}
  {\bibfnamefont {Z.}~\bibnamefont {Wang}}, \bibinfo {author} {\bibfnamefont
  {D.}~\bibnamefont {Zheng}}, \bibinfo {author} {\bibfnamefont
  {W.}~\bibnamefont {Hu}}, \bibinfo {author} {\bibfnamefont {X.}~\bibnamefont
  {Chen}}, \ and\ \bibinfo {author} {\bibfnamefont {W.}~\bibnamefont {Lu}},\
  }\bibfield  {title} {\enquote {\bibinfo {title} {Room-temperature
  single-photon detector based on single nanowire},}\ }\href {\doibase
  10.1021/acs.nanolett.8b01795} {\bibfield  {journal} {\bibinfo  {journal}
  {Nano Letters}\ }\textbf {\bibinfo {volume} {18}},\ \bibinfo {pages}
  {5439--5445} (\bibinfo {year} {2018}{\natexlab{b}})},\ \bibinfo {note} {pMID:
  30133292}\BibitemShut {NoStop}%
\bibitem [{\citenamefont {Chen}\ \emph {et~al.}(2018)\citenamefont {Chen},
  \citenamefont {Xia}, \citenamefont {Yang}, \citenamefont {Gong},
  \citenamefont {Wei}, \citenamefont {Wang}, \citenamefont {Tang},
  \citenamefont {Fang}, \citenamefont {Fang},\ and\ \citenamefont
  {Liao}}]{Chen2018}%
  \BibitemOpen
  \bibfield  {author} {\bibinfo {author} {\bibfnamefont {X.}~\bibnamefont
  {Chen}}, \bibinfo {author} {\bibfnamefont {N.}~\bibnamefont {Xia}}, \bibinfo
  {author} {\bibfnamefont {Z.}~\bibnamefont {Yang}}, \bibinfo {author}
  {\bibfnamefont {F.}~\bibnamefont {Gong}}, \bibinfo {author} {\bibfnamefont
  {Z.}~\bibnamefont {Wei}}, \bibinfo {author} {\bibfnamefont {D.}~\bibnamefont
  {Wang}}, \bibinfo {author} {\bibfnamefont {J.}~\bibnamefont {Tang}}, \bibinfo
  {author} {\bibfnamefont {X.}~\bibnamefont {Fang}}, \bibinfo {author}
  {\bibfnamefont {D.}~\bibnamefont {Fang}}, \ and\ \bibinfo {author}
  {\bibfnamefont {L.}~\bibnamefont {Liao}},\ }\bibfield  {title} {\enquote
  {\bibinfo {title} {Analysis of the influence and mechanism of sulfur
  passivation on the dark current of a single {GaAs} nanowire photodetector},}\
  }\href {\doibase 10.1088/1361-6528/aaa4d6} {\bibfield  {journal} {\bibinfo
  {journal} {Nanotechnology}\ }\textbf {\bibinfo {volume} {29}},\ \bibinfo
  {pages} {095201} (\bibinfo {year} {2018})}\BibitemShut {NoStop}%
\bibitem [{\citenamefont {Fang}\ \emph {et~al.}(2019)\citenamefont {Fang},
  \citenamefont {Armin}, \citenamefont {Meredith},\ and\ \citenamefont
  {Huang}}]{Fang2019}%
  \BibitemOpen
  \bibfield  {author} {\bibinfo {author} {\bibfnamefont {Y.}~\bibnamefont
  {Fang}}, \bibinfo {author} {\bibfnamefont {A.}~\bibnamefont {Armin}},
  \bibinfo {author} {\bibfnamefont {P.}~\bibnamefont {Meredith}}, \ and\
  \bibinfo {author} {\bibfnamefont {J.}~\bibnamefont {Huang}},\ }\bibfield
  {title} {\enquote {\bibinfo {title} {Accurate characterization of
  next-generation thin-film photodetectors},}\ }\href {\doibase
  10.1038/s41566-018-0288-z} {\bibfield  {journal} {\bibinfo  {journal} {Nature
  Photonics}\ }\textbf {\bibinfo {volume} {13}},\ \bibinfo {pages} {1--4}
  (\bibinfo {year} {2019})}\BibitemShut {NoStop}%
\bibitem [{\citenamefont {van Vliet}(1967)}]{vanVliet67}%
  \BibitemOpen
  \bibfield  {author} {\bibinfo {author} {\bibfnamefont {K.~M.}\ \bibnamefont
  {van Vliet}},\ }\bibfield  {title} {\enquote {\bibinfo {title} {Noise
  limitations in solid state photodetectors},}\ }\href {\doibase
  10.1364/AO.6.001145} {\bibfield  {journal} {\bibinfo  {journal} {Appl. Opt.}\
  }\textbf {\bibinfo {volume} {6}},\ \bibinfo {pages} {1145--1169} (\bibinfo
  {year} {1967})}\BibitemShut {NoStop}%
\bibitem [{\citenamefont {THORLABS}({\natexlab{a}})}]{ThorlabsNEP}%
  \BibitemOpen
  \bibfield  {author} {\bibinfo {author} {\bibnamefont {THORLABS}},\ }\bibfield
   {title} {\enquote {\bibinfo {title} {{NEP} – noise equivalent power},}\
  }\href
  {https://www.thorlabs.com/images/TabImages/Noise_Equivalent_Power_White_Paper.pdf}
  {\  ({\natexlab{a}})}\BibitemShut {NoStop}%
\bibitem [{\citenamefont {Murali}\ \emph {et~al.}(2019)\citenamefont {Murali},
  \citenamefont {Abraham}, \citenamefont {Das}, \citenamefont {Kallatt},\ and\
  \citenamefont {Majumdar}}]{Murali2019}%
  \BibitemOpen
  \bibfield  {author} {\bibinfo {author} {\bibfnamefont {K.}~\bibnamefont
  {Murali}}, \bibinfo {author} {\bibfnamefont {N.}~\bibnamefont {Abraham}},
  \bibinfo {author} {\bibfnamefont {S.}~\bibnamefont {Das}}, \bibinfo {author}
  {\bibfnamefont {S.}~\bibnamefont {Kallatt}}, \ and\ \bibinfo {author}
  {\bibfnamefont {K.}~\bibnamefont {Majumdar}},\ }\bibfield  {title} {\enquote
  {\bibinfo {title} {Highly sensitive, fast graphene photodetector with
  responsivity >106 A/W using a floating quantum well gate},}\ }\href {\doibase
  10.1021/acsami.9b06835} {\bibfield  {journal} {\bibinfo  {journal} {ACS
  Applied Materials \& Interfaces}\ }\textbf {\bibinfo {volume} {11}},\
  \bibinfo {pages} {30010--30018} (\bibinfo {year} {2019})}\BibitemShut
  {NoStop}%
\bibitem [{\citenamefont {Li}\ \emph {et~al.}(2016)\citenamefont {Li},
  \citenamefont {Zhu}, \citenamefont {Du}, \citenamefont {Lv}, \citenamefont
  {Zhang}, \citenamefont {Li}, \citenamefont {Yang}, \citenamefont {Yang},
  \citenamefont {Li}, \citenamefont {Wang}, \citenamefont {Zhu},\ and\
  \citenamefont {Fang}}]{Li2016}%
  \BibitemOpen
  \bibfield  {author} {\bibinfo {author} {\bibfnamefont {X.}~\bibnamefont
  {Li}}, \bibinfo {author} {\bibfnamefont {M.}~\bibnamefont {Zhu}}, \bibinfo
  {author} {\bibfnamefont {M.}~\bibnamefont {Du}}, \bibinfo {author}
  {\bibfnamefont {Z.}~\bibnamefont {Lv}}, \bibinfo {author} {\bibfnamefont
  {L.}~\bibnamefont {Zhang}}, \bibinfo {author} {\bibfnamefont
  {Y.}~\bibnamefont {Li}}, \bibinfo {author} {\bibfnamefont {Y.}~\bibnamefont
  {Yang}}, \bibinfo {author} {\bibfnamefont {T.}~\bibnamefont {Yang}}, \bibinfo
  {author} {\bibfnamefont {X.}~\bibnamefont {Li}}, \bibinfo {author}
  {\bibfnamefont {K.}~\bibnamefont {Wang}}, \bibinfo {author} {\bibfnamefont
  {H.}~\bibnamefont {Zhu}}, \ and\ \bibinfo {author} {\bibfnamefont
  {Y.}~\bibnamefont {Fang}},\ }\bibfield  {title} {\enquote {\bibinfo {title}
  {High detectivity graphene-silicon heterojunction photodetector},}\ }\href
  {\doibase 10.1002/smll.201502336} {\bibfield  {journal} {\bibinfo  {journal}
  {Small}\ }\textbf {\bibinfo {volume} {12}},\ \bibinfo {pages} {595--601}
  (\bibinfo {year} {2016})}\BibitemShut {NoStop}%
\bibitem [{\citenamefont {Luo}\ \emph {et~al.}(2015)\citenamefont {Luo},
  \citenamefont {Liang}, \citenamefont {Liu}, \citenamefont {Xie},
  \citenamefont {Lou},\ and\ \citenamefont {Shen}}]{Luo2015}%
  \BibitemOpen
  \bibfield  {author} {\bibinfo {author} {\bibfnamefont {T.}~\bibnamefont
  {Luo}}, \bibinfo {author} {\bibfnamefont {B.}~\bibnamefont {Liang}}, \bibinfo
  {author} {\bibfnamefont {Z.}~\bibnamefont {Liu}}, \bibinfo {author}
  {\bibfnamefont {X.}~\bibnamefont {Xie}}, \bibinfo {author} {\bibfnamefont
  {Z.}~\bibnamefont {Lou}}, \ and\ \bibinfo {author} {\bibfnamefont
  {G.}~\bibnamefont {Shen}},\ }\bibfield  {title} {\enquote {\bibinfo {title}
  {Single-gasb-nanowire-based room temperature photodetectors with broad
  spectral response},}\ }\href {\doibase 10.1007/s11434-014-0687-6} {\bibfield
  {journal} {\bibinfo  {journal} {Science Bulletin}\ }\textbf {\bibinfo
  {volume} {60}},\ \bibinfo {pages} {101 -- 108} (\bibinfo {year}
  {2015})}\BibitemShut {NoStop}%
\bibitem [{\citenamefont {Liu}\ \emph {et~al.}(2014)\citenamefont {Liu},
  \citenamefont {Chang}, \citenamefont {Norris},\ and\ \citenamefont
  {Zhong}}]{Liu2014}%
  \BibitemOpen
  \bibfield  {author} {\bibinfo {author} {\bibfnamefont {C.-H.}\ \bibnamefont
  {Liu}}, \bibinfo {author} {\bibfnamefont {Y.-C.}\ \bibnamefont {Chang}},
  \bibinfo {author} {\bibfnamefont {T.~B.}\ \bibnamefont {Norris}}, \ and\
  \bibinfo {author} {\bibfnamefont {Z.}~\bibnamefont {Zhong}},\ }\bibfield
  {title} {\enquote {\bibinfo {title} {Graphene photodetectors with
  ultra-broadband and high responsivity at room temperature},}\ }\href
  {\doibase 10.1038/nnano.2014.31} {\bibfield  {journal} {\bibinfo  {journal}
  {Nature Nanotechnology}\ }\textbf {\bibinfo {volume} {9}},\ \bibinfo {pages}
  {273 EP --} (\bibinfo {year} {2014})}\BibitemShut {NoStop}%
\bibitem [{\citenamefont {Lopez-Sanchez}\ \emph
  {et~al.}(2013{\natexlab{b}})\citenamefont {Lopez-Sanchez}, \citenamefont
  {Lembke}, \citenamefont {Kayci}, \citenamefont {Radenovic},\ and\
  \citenamefont {Kis}}]{Lopez-Sanchez2013}%
  \BibitemOpen
  \bibfield  {author} {\bibinfo {author} {\bibfnamefont {O.}~\bibnamefont
  {Lopez-Sanchez}}, \bibinfo {author} {\bibfnamefont {D.}~\bibnamefont
  {Lembke}}, \bibinfo {author} {\bibfnamefont {M.}~\bibnamefont {Kayci}},
  \bibinfo {author} {\bibfnamefont {A.}~\bibnamefont {Radenovic}}, \ and\
  \bibinfo {author} {\bibfnamefont {A.}~\bibnamefont {Kis}},\ }\bibfield
  {title} {\enquote {\bibinfo {title} {Ultrasensitive photodetectors based on
  monolayer mos2},}\ }\href {\doibase 10.1038/nnano.2013.100} {\bibfield
  {journal} {\bibinfo  {journal} {Nature Nanotechnology}\ }\textbf {\bibinfo
  {volume} {8}},\ \bibinfo {pages} {497 EP --} (\bibinfo {year}
  {2013}{\natexlab{b}})}\BibitemShut {NoStop}%
\bibitem [{\citenamefont {An}\ \emph {et~al.}(2013)\citenamefont {An},
  \citenamefont {Liu}, \citenamefont {Jung},\ and\ \citenamefont
  {Kar}}]{An2013}%
  \BibitemOpen
  \bibfield  {author} {\bibinfo {author} {\bibfnamefont {X.}~\bibnamefont
  {An}}, \bibinfo {author} {\bibfnamefont {F.}~\bibnamefont {Liu}}, \bibinfo
  {author} {\bibfnamefont {Y.~J.}\ \bibnamefont {Jung}}, \ and\ \bibinfo
  {author} {\bibfnamefont {S.}~\bibnamefont {Kar}},\ }\bibfield  {title}
  {\enquote {\bibinfo {title} {Tunable graphene-silicon heterojunctions for
  ultrasensitive photodetection},}\ }\href {\doibase 10.1021/nl303682j}
  {\bibfield  {journal} {\bibinfo  {journal} {Nano Letters}\ }\textbf {\bibinfo
  {volume} {13}},\ \bibinfo {pages} {909--916} (\bibinfo {year}
  {2013})}\BibitemShut {NoStop}%
\bibitem [{\citenamefont {Konstantatos}\ \emph {et~al.}(2007)\citenamefont
  {Konstantatos}, \citenamefont {Clifford}, \citenamefont {Levina},\ and\
  \citenamefont {Sargent}}]{Konstantatos2007}%
  \BibitemOpen
  \bibfield  {author} {\bibinfo {author} {\bibfnamefont {G.}~\bibnamefont
  {Konstantatos}}, \bibinfo {author} {\bibfnamefont {J.}~\bibnamefont
  {Clifford}}, \bibinfo {author} {\bibfnamefont {L.}~\bibnamefont {Levina}}, \
  and\ \bibinfo {author} {\bibfnamefont {E.~H.}\ \bibnamefont {Sargent}},\
  }\bibfield  {title} {\enquote {\bibinfo {title} {Sensitive solution-processed
  visible-wavelength photodetectors},}\ }\href {\doibase
  10.1038/nphoton.2007.147} {\bibfield  {journal} {\bibinfo  {journal} {Nature
  Photonics}\ }\textbf {\bibinfo {volume} {1}},\ \bibinfo {pages} {531 EP --}
  (\bibinfo {year} {2007})}\BibitemShut {NoStop}%
\bibitem [{\citenamefont {Kim}\ \emph {et~al.}(2015)\citenamefont {Kim},
  \citenamefont {Adinolfi}, \citenamefont {Sutherland}, \citenamefont {Voznyy},
  \citenamefont {Kwon}, \citenamefont {Kim}, \citenamefont {Kim}, \citenamefont
  {Ihee}, \citenamefont {Kemp}, \citenamefont {Adachi}, \citenamefont {Yuan},
  \citenamefont {Kramer}, \citenamefont {Zhitomirsky}, \citenamefont
  {Hoogland},\ and\ \citenamefont {Sargent}}]{Kim2015}%
  \BibitemOpen
  \bibfield  {author} {\bibinfo {author} {\bibfnamefont {J.~Y.}\ \bibnamefont
  {Kim}}, \bibinfo {author} {\bibfnamefont {V.}~\bibnamefont {Adinolfi}},
  \bibinfo {author} {\bibfnamefont {B.~R.}\ \bibnamefont {Sutherland}},
  \bibinfo {author} {\bibfnamefont {O.}~\bibnamefont {Voznyy}}, \bibinfo
  {author} {\bibfnamefont {S.~J.}\ \bibnamefont {Kwon}}, \bibinfo {author}
  {\bibfnamefont {T.~W.}\ \bibnamefont {Kim}}, \bibinfo {author} {\bibfnamefont
  {J.}~\bibnamefont {Kim}}, \bibinfo {author} {\bibfnamefont {H.}~\bibnamefont
  {Ihee}}, \bibinfo {author} {\bibfnamefont {K.}~\bibnamefont {Kemp}}, \bibinfo
  {author} {\bibfnamefont {M.}~\bibnamefont {Adachi}}, \bibinfo {author}
  {\bibfnamefont {M.}~\bibnamefont {Yuan}}, \bibinfo {author} {\bibfnamefont
  {I.}~\bibnamefont {Kramer}}, \bibinfo {author} {\bibfnamefont
  {D.}~\bibnamefont {Zhitomirsky}}, \bibinfo {author} {\bibfnamefont
  {S.}~\bibnamefont {Hoogland}}, \ and\ \bibinfo {author} {\bibfnamefont
  {E.~H.}\ \bibnamefont {Sargent}},\ }\bibfield  {title} {\enquote {\bibinfo
  {title} {Single-step fabrication of quantum funnels via centrifugal colloidal
  casting of nanoparticle films},}\ }\href {\doibase 10.1038/ncomms8772}
  {\bibfield  {journal} {\bibinfo  {journal} {Nature Communications}\ }\textbf
  {\bibinfo {volume} {6}},\ \bibinfo {pages} {7772 EP --} (\bibinfo {year}
  {2015})}\BibitemShut {NoStop}%
\bibitem [{\citenamefont {Shen}\ \emph {et~al.}(2016)\citenamefont {Shen},
  \citenamefont {Zhang}, \citenamefont {Bai}, \citenamefont {Zheng},
  \citenamefont {Wang},\ and\ \citenamefont {Huang}}]{Shen2016}%
  \BibitemOpen
  \bibfield  {author} {\bibinfo {author} {\bibfnamefont {L.}~\bibnamefont
  {Shen}}, \bibinfo {author} {\bibfnamefont {Y.}~\bibnamefont {Zhang}},
  \bibinfo {author} {\bibfnamefont {Y.}~\bibnamefont {Bai}}, \bibinfo {author}
  {\bibfnamefont {X.}~\bibnamefont {Zheng}}, \bibinfo {author} {\bibfnamefont
  {Q.}~\bibnamefont {Wang}}, \ and\ \bibinfo {author} {\bibfnamefont
  {J.}~\bibnamefont {Huang}},\ }\bibfield  {title} {\enquote {\bibinfo {title}
  {A filterless{,} visible-blind{,} narrow-band{,} and near-infrared
  photodetector with a gain},}\ }\href {\doibase 10.1039/C6NR02902G} {\bibfield
   {journal} {\bibinfo  {journal} {Nanoscale}\ }\textbf {\bibinfo {volume}
  {8}},\ \bibinfo {pages} {12990--12997} (\bibinfo {year} {2016})}\BibitemShut
  {NoStop}%
\bibitem [{\citenamefont {{Casaluci}}\ \emph {et~al.}(2016)\citenamefont
  {{Casaluci}}, \citenamefont {{Cinà}}, \citenamefont {{Matteocci}},
  \citenamefont {{Lugli}},\ and\ \citenamefont {{Di Carlo}}}]{Casaluci2016}%
  \BibitemOpen
  \bibfield  {author} {\bibinfo {author} {\bibfnamefont {S.}~\bibnamefont
  {{Casaluci}}}, \bibinfo {author} {\bibfnamefont {L.}~\bibnamefont {{Cinà}}},
  \bibinfo {author} {\bibfnamefont {F.}~\bibnamefont {{Matteocci}}}, \bibinfo
  {author} {\bibfnamefont {P.}~\bibnamefont {{Lugli}}}, \ and\ \bibinfo
  {author} {\bibfnamefont {A.}~\bibnamefont {{Di Carlo}}},\ }\bibfield  {title}
  {\enquote {\bibinfo {title} {Fabrication and characterization of mesoscopic
  perovskite photodiodes},}\ }\href {\doibase 10.1109/TNANO.2016.2517239}
  {\bibfield  {journal} {\bibinfo  {journal} {IEEE Transactions on
  Nanotechnology}\ }\textbf {\bibinfo {volume} {15}},\ \bibinfo {pages}
  {255--260} (\bibinfo {year} {2016})}\BibitemShut {NoStop}%
\bibitem [{\citenamefont {Wang}\ \emph {et~al.}(2018)\citenamefont {Wang},
  \citenamefont {Du}, \citenamefont {Zhang}, \citenamefont {Miao},
  \citenamefont {Fang},\ and\ \citenamefont {Zhang}}]{Wang2018}%
  \BibitemOpen
  \bibfield  {author} {\bibinfo {author} {\bibfnamefont {W.}~\bibnamefont
  {Wang}}, \bibinfo {author} {\bibfnamefont {M.}~\bibnamefont {Du}}, \bibinfo
  {author} {\bibfnamefont {M.}~\bibnamefont {Zhang}}, \bibinfo {author}
  {\bibfnamefont {J.}~\bibnamefont {Miao}}, \bibinfo {author} {\bibfnamefont
  {Y.}~\bibnamefont {Fang}}, \ and\ \bibinfo {author} {\bibfnamefont
  {F.}~\bibnamefont {Zhang}},\ }\bibfield  {title} {\enquote {\bibinfo {title}
  {Organic photodetectors with gain and broadband/narrowband response under
  top/bottom illumination conditions},}\ }\href {\doibase
  10.1002/adom.201800249} {\bibfield  {journal} {\bibinfo  {journal} {Advanced
  Optical Materials}\ }\textbf {\bibinfo {volume} {6}},\ \bibinfo {pages}
  {1800249} (\bibinfo {year} {2018})}\BibitemShut {NoStop}%
\bibitem [{\citenamefont {Sutherland}\ \emph {et~al.}(2015)\citenamefont
  {Sutherland}, \citenamefont {Johnston}, \citenamefont {Ip}, \citenamefont
  {Xu}, \citenamefont {Adinolfi}, \citenamefont {Kanjanaboos},\ and\
  \citenamefont {Sargent}}]{Sutherland2015}%
  \BibitemOpen
  \bibfield  {author} {\bibinfo {author} {\bibfnamefont {B.~R.}\ \bibnamefont
  {Sutherland}}, \bibinfo {author} {\bibfnamefont {A.~K.}\ \bibnamefont
  {Johnston}}, \bibinfo {author} {\bibfnamefont {A.~H.}\ \bibnamefont {Ip}},
  \bibinfo {author} {\bibfnamefont {J.}~\bibnamefont {Xu}}, \bibinfo {author}
  {\bibfnamefont {V.}~\bibnamefont {Adinolfi}}, \bibinfo {author}
  {\bibfnamefont {P.}~\bibnamefont {Kanjanaboos}}, \ and\ \bibinfo {author}
  {\bibfnamefont {E.~H.}\ \bibnamefont {Sargent}},\ }\bibfield  {title}
  {\enquote {\bibinfo {title} {Sensitive, fast, and stable perovskite
  photodetectors exploiting interface engineering},}\ }\href {\doibase
  10.1021/acsphotonics.5b00164} {\bibfield  {journal} {\bibinfo  {journal} {ACS
  Photonics}\ }\textbf {\bibinfo {volume} {2}},\ \bibinfo {pages} {1117--1123}
  (\bibinfo {year} {2015})}\BibitemShut {NoStop}%
\bibitem [{\citenamefont {Konstantatos}\ \emph {et~al.}(2006)\citenamefont
  {Konstantatos}, \citenamefont {Howard}, \citenamefont {Fischer},
  \citenamefont {Hoogland}, \citenamefont {Clifford}, \citenamefont {Klem},
  \citenamefont {Levina},\ and\ \citenamefont {Sargent}}]{Konstantatos2006}%
  \BibitemOpen
  \bibfield  {author} {\bibinfo {author} {\bibfnamefont {G.}~\bibnamefont
  {Konstantatos}}, \bibinfo {author} {\bibfnamefont {I.}~\bibnamefont
  {Howard}}, \bibinfo {author} {\bibfnamefont {A.}~\bibnamefont {Fischer}},
  \bibinfo {author} {\bibfnamefont {S.}~\bibnamefont {Hoogland}}, \bibinfo
  {author} {\bibfnamefont {J.}~\bibnamefont {Clifford}}, \bibinfo {author}
  {\bibfnamefont {E.}~\bibnamefont {Klem}}, \bibinfo {author} {\bibfnamefont
  {L.}~\bibnamefont {Levina}}, \ and\ \bibinfo {author} {\bibfnamefont {E.~H.}\
  \bibnamefont {Sargent}},\ }\bibfield  {title} {\enquote {\bibinfo {title}
  {Ultrasensitive solution-cast quantum dot photodetectors},}\ }\href {\doibase
  10.1038/nature04855} {\bibfield  {journal} {\bibinfo  {journal} {Nature}\
  }\textbf {\bibinfo {volume} {442}},\ \bibinfo {pages} {180 EP --} (\bibinfo
  {year} {2006})}\BibitemShut {NoStop}%
\bibitem [{\citenamefont {THORLABS}({\natexlab{b}})}]{FDS015}%
  \BibitemOpen
  \bibfield  {author} {\bibinfo {author} {\bibnamefont {THORLABS}},\ }\bibfield
   {title} {\enquote {\bibinfo {title} {{FDS015}},}\ }\href
  {https://www.thorlabs.com/thorproduct.cfm?partnumber=FDS015} {\
  ({\natexlab{b}})}\BibitemShut {NoStop}%
\bibitem [{\citenamefont {THORLABS}({\natexlab{c}})}]{FGAP71}%
  \BibitemOpen
  \bibfield  {author} {\bibinfo {author} {\bibnamefont {THORLABS}},\ }\bibfield
   {title} {\enquote {\bibinfo {title} {{FGAP71}},}\ }\href
  {https://www.thorlabs.com/thorproduct.cfm?partnumber=FGAP71} {\
  ({\natexlab{c}})}\BibitemShut {NoStop}%
\bibitem [{\citenamefont {Kufer}\ \emph {et~al.}(2015)\citenamefont {Kufer},
  \citenamefont {Nikitskiy}, \citenamefont {Lasanta}, \citenamefont
  {Navickaite}, \citenamefont {Koppens},\ and\ \citenamefont
  {Konstantatos}}]{Kufer2015}%
  \BibitemOpen
  \bibfield  {author} {\bibinfo {author} {\bibfnamefont {D.}~\bibnamefont
  {Kufer}}, \bibinfo {author} {\bibfnamefont {I.}~\bibnamefont {Nikitskiy}},
  \bibinfo {author} {\bibfnamefont {T.}~\bibnamefont {Lasanta}}, \bibinfo
  {author} {\bibfnamefont {G.}~\bibnamefont {Navickaite}}, \bibinfo {author}
  {\bibfnamefont {F.~H.~L.}\ \bibnamefont {Koppens}}, \ and\ \bibinfo {author}
  {\bibfnamefont {G.}~\bibnamefont {Konstantatos}},\ }\bibfield  {title}
  {\enquote {\bibinfo {title} {Hybrid 2d–0d mos2–pbs quantum dot
  photodetectors},}\ }\href {\doibase 10.1002/adma.201402471} {\bibfield
  {journal} {\bibinfo  {journal} {Advanced Materials}\ }\textbf {\bibinfo
  {volume} {27}},\ \bibinfo {pages} {176--180} (\bibinfo {year}
  {2015})}\BibitemShut {NoStop}%
\bibitem [{\citenamefont {Huo}\ and\ \citenamefont
  {Konstantatos}(2017)}]{Huo2017}%
  \BibitemOpen
  \bibfield  {author} {\bibinfo {author} {\bibfnamefont {N.}~\bibnamefont
  {Huo}}\ and\ \bibinfo {author} {\bibfnamefont {G.}~\bibnamefont
  {Konstantatos}},\ }\bibfield  {title} {\enquote {\bibinfo {title}
  {Ultrasensitive all-2d mos2 phototransistors enabled by an out-of-plane mos2
  pn homojunction},}\ }\href {\doibase 10.1038/s41467-017-00722-1} {\bibfield
  {journal} {\bibinfo  {journal} {Nature Communications}\ }\textbf {\bibinfo
  {volume} {8}},\ \bibinfo {pages} {572} (\bibinfo {year} {2017})}\BibitemShut
  {NoStop}%
\bibitem [{\citenamefont {Hwang}\ \emph {et~al.}(2016)\citenamefont {Hwang},
  \citenamefont {Lee}, \citenamefont {Lee}, \citenamefont {Lee}, \citenamefont
  {Shokouh}, \citenamefont {Kyhm}, \citenamefont {Lee}, \citenamefont {Kim},
  \citenamefont {Yoo}, \citenamefont {Nam}, \citenamefont {Son}, \citenamefont
  {Ju}, \citenamefont {Park}, \citenamefont {Song}, \citenamefont {Choi},\ and\
  \citenamefont {Im}}]{Hwang2016}%
  \BibitemOpen
  \bibfield  {author} {\bibinfo {author} {\bibfnamefont {D.~K.}\ \bibnamefont
  {Hwang}}, \bibinfo {author} {\bibfnamefont {Y.~T.}\ \bibnamefont {Lee}},
  \bibinfo {author} {\bibfnamefont {H.~S.}\ \bibnamefont {Lee}}, \bibinfo
  {author} {\bibfnamefont {Y.~J.}\ \bibnamefont {Lee}}, \bibinfo {author}
  {\bibfnamefont {S.~H.}\ \bibnamefont {Shokouh}}, \bibinfo {author}
  {\bibfnamefont {J.-h.}\ \bibnamefont {Kyhm}}, \bibinfo {author}
  {\bibfnamefont {J.}~\bibnamefont {Lee}}, \bibinfo {author} {\bibfnamefont
  {H.~H.}\ \bibnamefont {Kim}}, \bibinfo {author} {\bibfnamefont {T.-H.}\
  \bibnamefont {Yoo}}, \bibinfo {author} {\bibfnamefont {S.~H.}\ \bibnamefont
  {Nam}}, \bibinfo {author} {\bibfnamefont {D.~I.}\ \bibnamefont {Son}},
  \bibinfo {author} {\bibfnamefont {B.-K.}\ \bibnamefont {Ju}}, \bibinfo
  {author} {\bibfnamefont {M.-C.}\ \bibnamefont {Park}}, \bibinfo {author}
  {\bibfnamefont {J.~D.}\ \bibnamefont {Song}}, \bibinfo {author}
  {\bibfnamefont {W.~K.}\ \bibnamefont {Choi}}, \ and\ \bibinfo {author}
  {\bibfnamefont {S.}~\bibnamefont {Im}},\ }\bibfield  {title} {\enquote
  {\bibinfo {title} {Ultrasensitive pbs quantum-dot-sensitized ingazno hybrid
  photoinverter for near-infrared detection and imaging with high photogain},}\
  }\href {\doibase 10.1038/am.2015.137} {\bibfield  {journal} {\bibinfo
  {journal} {Npg Asia Materials}\ }\textbf {\bibinfo {volume} {8}},\ \bibinfo
  {pages} {e233 EP --} (\bibinfo {year} {2016})}\BibitemShut {NoStop}%
\bibitem [{\citenamefont {{Murali}}\ and\ \citenamefont
  {{Majumdar}}(2018)}]{Murali2018}%
  \BibitemOpen
  \bibfield  {author} {\bibinfo {author} {\bibfnamefont {K.}~\bibnamefont
  {{Murali}}}\ and\ \bibinfo {author} {\bibfnamefont {K.}~\bibnamefont
  {{Majumdar}}},\ }\bibfield  {title} {\enquote {\bibinfo {title}
  {Self-powered, highly sensitive, high-speed photodetection using
  ITO/WSe2/SnSe2 vertical heterojunction},}\ }\href {\doibase
  10.1109/TED.2018.2864250} {\bibfield  {journal} {\bibinfo  {journal} {IEEE
  Transactions on Electron Devices}\ }\textbf {\bibinfo {volume} {65}},\
  \bibinfo {pages} {4141} (\bibinfo {year} {2018})}\BibitemShut {NoStop}%
\bibitem [{\citenamefont {Kallatt}, \citenamefont {Nair},\ and\ \citenamefont
  {Majumdar}(2018)}]{Kallat2018}%
  \BibitemOpen
  \bibfield  {author} {\bibinfo {author} {\bibfnamefont {S.}~\bibnamefont
  {Kallatt}}, \bibinfo {author} {\bibfnamefont {S.}~\bibnamefont {Nair}}, \
  and\ \bibinfo {author} {\bibfnamefont {K.}~\bibnamefont {Majumdar}},\
  }\bibfield  {title} {\enquote {\bibinfo {title} {Asymmetrically encapsulated
  vertical ITO/MoS2/Cu2O photodetector with ultrahigh sensitivity},}\ }\href
  {\doibase 10.1002/smll.201702066} {\bibfield  {journal} {\bibinfo  {journal}
  {Small}\ }\textbf {\bibinfo {volume} {14}},\ \bibinfo {pages} {1702066}
  (\bibinfo {year} {2018})}\BibitemShut {NoStop}%
\bibitem [{\citenamefont {{Dhyani}}\ \emph {et~al.}(2020)\citenamefont
  {{Dhyani}}, \citenamefont {{Ahmad}}, \citenamefont {{Kumar}},\ and\
  \citenamefont {{Das}}}]{Dhyani20}%
  \BibitemOpen
  \bibfield  {author} {\bibinfo {author} {\bibfnamefont {V.}~\bibnamefont
  {{Dhyani}}}, \bibinfo {author} {\bibfnamefont {G.}~\bibnamefont {{Ahmad}}},
  \bibinfo {author} {\bibfnamefont {N.}~\bibnamefont {{Kumar}}}, \ and\
  \bibinfo {author} {\bibfnamefont {S.}~\bibnamefont {{Das}}},\ }\bibfield
  {title} {\enquote {\bibinfo {title} {Size-dependent photoresponse of
  germanium nanocrystals-metal oxide semiconductor photodetector},}\
  }\href@noop {} {\bibfield  {journal} {\bibinfo  {journal} {IEEE Transactions
  on Electron Devices}\ }\textbf {\bibinfo {volume} {67}},\ \bibinfo {pages}
  {558--565} (\bibinfo {year} {2020})}\BibitemShut {NoStop}%
\bibitem [{\citenamefont {Huang}\ \emph {et~al.}(2020)\citenamefont {Huang},
  \citenamefont {Lee}, \citenamefont {Vollbrecht}, \citenamefont {Brus},
  \citenamefont {Dixon}, \citenamefont {Cao}, \citenamefont {Zhu},
  \citenamefont {Du}, \citenamefont {Wang}, \citenamefont {Cho}, \citenamefont
  {Bazan},\ and\ \citenamefont {Nguyen}}]{Huang20}%
  \BibitemOpen
  \bibfield  {author} {\bibinfo {author} {\bibfnamefont {J.}~\bibnamefont
  {Huang}}, \bibinfo {author} {\bibfnamefont {J.}~\bibnamefont {Lee}}, \bibinfo
  {author} {\bibfnamefont {J.}~\bibnamefont {Vollbrecht}}, \bibinfo {author}
  {\bibfnamefont {V.~V.}\ \bibnamefont {Brus}}, \bibinfo {author}
  {\bibfnamefont {A.~L.}\ \bibnamefont {Dixon}}, \bibinfo {author}
  {\bibfnamefont {D.~X.}\ \bibnamefont {Cao}}, \bibinfo {author} {\bibfnamefont
  {Z.}~\bibnamefont {Zhu}}, \bibinfo {author} {\bibfnamefont {Z.}~\bibnamefont
  {Du}}, \bibinfo {author} {\bibfnamefont {H.}~\bibnamefont {Wang}}, \bibinfo
  {author} {\bibfnamefont {K.}~\bibnamefont {Cho}}, \bibinfo {author}
  {\bibfnamefont {G.~C.}\ \bibnamefont {Bazan}}, \ and\ \bibinfo {author}
  {\bibfnamefont {T.-Q.}\ \bibnamefont {Nguyen}},\ }\bibfield  {title}
  {\enquote {\bibinfo {title} {A high-performance solution-processed organic
  photodetector for near-infrared sensing},}\ }\href {\doibase
  10.1002/adma.201906027} {\bibfield  {journal} {\bibinfo  {journal} {Advanced
  Materials}\ }\textbf {\bibinfo {volume} {32}},\ \bibinfo {pages} {1906027}
  (\bibinfo {year} {2020})},\ \Eprint
  {http://arxiv.org/abs/https://onlinelibrary.wiley.com/doi/pdf/10.1002/adma.201906027}
  {https://onlinelibrary.wiley.com/doi/pdf/10.1002/adma.201906027} \BibitemShut
  {NoStop}%
\bibitem [{\citenamefont {Li}\ \emph {et~al.}(2020)\citenamefont {Li},
  \citenamefont {Xia}, \citenamefont {Liu}, \citenamefont {Zhang},
  \citenamefont {Teng}, \citenamefont {Zhang}, \citenamefont {Liu},
  \citenamefont {Zhao}, \citenamefont {Zhao}, \citenamefont {Li}, \citenamefont
  {Xing}, \citenamefont {Kang},\ and\ \citenamefont {Wei}}]{Li20}%
  \BibitemOpen
  \bibfield  {author} {\bibinfo {author} {\bibfnamefont {J.}~\bibnamefont
  {Li}}, \bibinfo {author} {\bibfnamefont {J.}~\bibnamefont {Xia}}, \bibinfo
  {author} {\bibfnamefont {Y.}~\bibnamefont {Liu}}, \bibinfo {author}
  {\bibfnamefont {S.}~\bibnamefont {Zhang}}, \bibinfo {author} {\bibfnamefont
  {C.}~\bibnamefont {Teng}}, \bibinfo {author} {\bibfnamefont {X.}~\bibnamefont
  {Zhang}}, \bibinfo {author} {\bibfnamefont {B.}~\bibnamefont {Liu}}, \bibinfo
  {author} {\bibfnamefont {S.}~\bibnamefont {Zhao}}, \bibinfo {author}
  {\bibfnamefont {S.}~\bibnamefont {Zhao}}, \bibinfo {author} {\bibfnamefont
  {B.}~\bibnamefont {Li}}, \bibinfo {author} {\bibfnamefont {G.}~\bibnamefont
  {Xing}}, \bibinfo {author} {\bibfnamefont {F.}~\bibnamefont {Kang}}, \ and\
  \bibinfo {author} {\bibfnamefont {G.}~\bibnamefont {Wei}},\ }\bibfield
  {title} {\enquote {\bibinfo {title} {Ultrasensitive organic-modulated cspbbr3
  quantum dot photodetectors via fast interfacial charge transfer},}\ }\href
  {\doibase 10.1002/admi.201901741} {\bibfield  {journal} {\bibinfo  {journal}
  {Advanced Materials Interfaces}\ }\textbf {\bibinfo {volume} {7}},\ \bibinfo
  {pages} {1901741} (\bibinfo {year} {2020})},\ \Eprint
  {http://arxiv.org/abs/https://onlinelibrary.wiley.com/doi/pdf/10.1002/admi.201901741}
  {https://onlinelibrary.wiley.com/doi/pdf/10.1002/admi.201901741} \BibitemShut
  {NoStop}%
\bibitem [{\citenamefont {Bao}\ \emph {et~al.}(2018)\citenamefont {Bao},
  \citenamefont {Yang}, \citenamefont {Bai}, \citenamefont {Xu}, \citenamefont
  {Yan}, \citenamefont {Xu}, \citenamefont {Liu}, \citenamefont {Zhang},\ and\
  \citenamefont {Gao}}]{Bao18}%
  \BibitemOpen
  \bibfield  {author} {\bibinfo {author} {\bibfnamefont {C.}~\bibnamefont
  {Bao}}, \bibinfo {author} {\bibfnamefont {J.}~\bibnamefont {Yang}}, \bibinfo
  {author} {\bibfnamefont {S.}~\bibnamefont {Bai}}, \bibinfo {author}
  {\bibfnamefont {W.}~\bibnamefont {Xu}}, \bibinfo {author} {\bibfnamefont
  {Z.}~\bibnamefont {Yan}}, \bibinfo {author} {\bibfnamefont {Q.}~\bibnamefont
  {Xu}}, \bibinfo {author} {\bibfnamefont {J.}~\bibnamefont {Liu}}, \bibinfo
  {author} {\bibfnamefont {W.}~\bibnamefont {Zhang}}, \ and\ \bibinfo {author}
  {\bibfnamefont {F.}~\bibnamefont {Gao}},\ }\bibfield  {title} {\enquote
  {\bibinfo {title} {High performance and stable all-inorganic metal halide
  perovskite-based photodetectors for optical communication applications},}\
  }\href {\doibase 10.1002/adma.201803422} {\bibfield  {journal} {\bibinfo
  {journal} {Advanced Materials}\ }\textbf {\bibinfo {volume} {30}},\ \bibinfo
  {pages} {1803422} (\bibinfo {year} {2018})},\ \Eprint
  {http://arxiv.org/abs/https://onlinelibrary.wiley.com/doi/pdf/10.1002/adma.201803422}
  {https://onlinelibrary.wiley.com/doi/pdf/10.1002/adma.201803422} \BibitemShut
  {NoStop}%
\bibitem [{\citenamefont {Cakmakyapan}\ \emph {et~al.}(2018)\citenamefont
  {Cakmakyapan}, \citenamefont {Lu}, \citenamefont {Navabi},\ and\
  \citenamefont {Jarrahi}}]{Cakmakyapan2018}%
  \BibitemOpen
  \bibfield  {author} {\bibinfo {author} {\bibfnamefont {S.}~\bibnamefont
  {Cakmakyapan}}, \bibinfo {author} {\bibfnamefont {P.~K.}\ \bibnamefont {Lu}},
  \bibinfo {author} {\bibfnamefont {A.}~\bibnamefont {Navabi}}, \ and\ \bibinfo
  {author} {\bibfnamefont {M.}~\bibnamefont {Jarrahi}},\ }\bibfield  {title}
  {\enquote {\bibinfo {title} {Gold-patched graphene nano-stripes for
  high-responsivity and ultrafast photodetection from the visible to infrared
  regime},}\ }\href {\doibase 10.1038/s41377-018-0020-2} {\bibfield  {journal}
  {\bibinfo  {journal} {Light: Science \& Applications}\ }\textbf {\bibinfo
  {volume} {7}},\ \bibinfo {pages} {20} (\bibinfo {year} {2018})}\BibitemShut
  {NoStop}%
\bibitem [{\citenamefont {Yu}\ \emph {et~al.}(2016)\citenamefont {Yu},
  \citenamefont {Wang}, \citenamefont {Xu}, \citenamefont {Ma}, \citenamefont
  {Pi},\ and\ \citenamefont {Yang}}]{Yu16}%
  \BibitemOpen
  \bibfield  {author} {\bibinfo {author} {\bibfnamefont {T.}~\bibnamefont
  {Yu}}, \bibinfo {author} {\bibfnamefont {F.}~\bibnamefont {Wang}}, \bibinfo
  {author} {\bibfnamefont {Y.}~\bibnamefont {Xu}}, \bibinfo {author}
  {\bibfnamefont {L.}~\bibnamefont {Ma}}, \bibinfo {author} {\bibfnamefont
  {X.}~\bibnamefont {Pi}}, \ and\ \bibinfo {author} {\bibfnamefont
  {D.}~\bibnamefont {Yang}},\ }\bibfield  {title} {\enquote {\bibinfo {title}
  {Graphene coupled with silicon quantum dots for high-performance
  bulk-silicon-based schottky-junction photodetectors},}\ }\href {\doibase
  10.1002/adma.201506140} {\bibfield  {journal} {\bibinfo  {journal} {Advanced
  Materials}\ }\textbf {\bibinfo {volume} {28}},\ \bibinfo {pages} {4912--4919}
  (\bibinfo {year} {2016})},\ \Eprint
  {http://arxiv.org/abs/https://onlinelibrary.wiley.com/doi/pdf/10.1002/adma.201506140}
  {https://onlinelibrary.wiley.com/doi/pdf/10.1002/adma.201506140} \BibitemShut
  {NoStop}%
\bibitem [{\citenamefont {Xu}\ \emph {et~al.}(2017)\citenamefont {Xu},
  \citenamefont {Ali}, \citenamefont {Shehzad}, \citenamefont {Meng},
  \citenamefont {Xu}, \citenamefont {Zhang}, \citenamefont {Wang},
  \citenamefont {Jin}, \citenamefont {Wang}, \citenamefont {Guo}, \citenamefont
  {Yang}, \citenamefont {Yu}, \citenamefont {Liu}, \citenamefont {He},
  \citenamefont {Duan}, \citenamefont {Wang}, \citenamefont {Tan},
  \citenamefont {Hu}, \citenamefont {Lu},\ and\ \citenamefont {Hasan}}]{Xu17}%
  \BibitemOpen
  \bibfield  {author} {\bibinfo {author} {\bibfnamefont {Y.}~\bibnamefont
  {Xu}}, \bibinfo {author} {\bibfnamefont {A.}~\bibnamefont {Ali}}, \bibinfo
  {author} {\bibfnamefont {K.}~\bibnamefont {Shehzad}}, \bibinfo {author}
  {\bibfnamefont {N.}~\bibnamefont {Meng}}, \bibinfo {author} {\bibfnamefont
  {M.}~\bibnamefont {Xu}}, \bibinfo {author} {\bibfnamefont {Y.}~\bibnamefont
  {Zhang}}, \bibinfo {author} {\bibfnamefont {X.}~\bibnamefont {Wang}},
  \bibinfo {author} {\bibfnamefont {C.}~\bibnamefont {Jin}}, \bibinfo {author}
  {\bibfnamefont {H.}~\bibnamefont {Wang}}, \bibinfo {author} {\bibfnamefont
  {Y.}~\bibnamefont {Guo}}, \bibinfo {author} {\bibfnamefont {Z.}~\bibnamefont
  {Yang}}, \bibinfo {author} {\bibfnamefont {B.}~\bibnamefont {Yu}}, \bibinfo
  {author} {\bibfnamefont {Y.}~\bibnamefont {Liu}}, \bibinfo {author}
  {\bibfnamefont {Q.}~\bibnamefont {He}}, \bibinfo {author} {\bibfnamefont
  {X.}~\bibnamefont {Duan}}, \bibinfo {author} {\bibfnamefont {X.}~\bibnamefont
  {Wang}}, \bibinfo {author} {\bibfnamefont {P.-H.}\ \bibnamefont {Tan}},
  \bibinfo {author} {\bibfnamefont {W.}~\bibnamefont {Hu}}, \bibinfo {author}
  {\bibfnamefont {H.}~\bibnamefont {Lu}}, \ and\ \bibinfo {author}
  {\bibfnamefont {T.}~\bibnamefont {Hasan}},\ }\bibfield  {title} {\enquote
  {\bibinfo {title} {Solvent-based soft-patterning of graphene lateral
  heterostructures for broadband high-speed metal–semiconductor–metal
  photodetectors},}\ }\href {\doibase 10.1002/admt.201600241} {\bibfield
  {journal} {\bibinfo  {journal} {Advanced Materials Technologies}\ }\textbf
  {\bibinfo {volume} {2}},\ \bibinfo {pages} {1600241} (\bibinfo {year}
  {2017})},\ \Eprint
  {http://arxiv.org/abs/https://onlinelibrary.wiley.com/doi/pdf/10.1002/admt.201600241}
  {https://onlinelibrary.wiley.com/doi/pdf/10.1002/admt.201600241} \BibitemShut
  {NoStop}%
\bibitem [{\citenamefont {Kim}\ \emph {et~al.}(2014)\citenamefont {Kim},
  \citenamefont {Kim}, \citenamefont {Shin}, \citenamefont {Kang},
  \citenamefont {Kim}, \citenamefont {Jang}, \citenamefont {Joo}, \citenamefont
  {Lee}, \citenamefont {Kim}, \citenamefont {Choi},\ and\ \citenamefont
  {Hwang}}]{Kim2014}%
  \BibitemOpen
  \bibfield  {author} {\bibinfo {author} {\bibfnamefont {C.~O.}\ \bibnamefont
  {Kim}}, \bibinfo {author} {\bibfnamefont {S.}~\bibnamefont {Kim}}, \bibinfo
  {author} {\bibfnamefont {D.~H.}\ \bibnamefont {Shin}}, \bibinfo {author}
  {\bibfnamefont {S.~S.}\ \bibnamefont {Kang}}, \bibinfo {author}
  {\bibfnamefont {J.~M.}\ \bibnamefont {Kim}}, \bibinfo {author} {\bibfnamefont
  {C.~W.}\ \bibnamefont {Jang}}, \bibinfo {author} {\bibfnamefont {S.~S.}\
  \bibnamefont {Joo}}, \bibinfo {author} {\bibfnamefont {J.~S.}\ \bibnamefont
  {Lee}}, \bibinfo {author} {\bibfnamefont {J.~H.}\ \bibnamefont {Kim}},
  \bibinfo {author} {\bibfnamefont {S.-H.}\ \bibnamefont {Choi}}, \ and\
  \bibinfo {author} {\bibfnamefont {E.}~\bibnamefont {Hwang}},\ }\bibfield
  {title} {\enquote {\bibinfo {title} {High photoresponsivity in an
  all-graphene p-n vertical junction photodetector},}\ }\href {\doibase
  10.1038/ncomms4249} {\bibfield  {journal} {\bibinfo  {journal} {Nature
  Communications}\ }\textbf {\bibinfo {volume} {5}},\ \bibinfo {pages} {3249}
  (\bibinfo {year} {2014})}\BibitemShut {NoStop}%
\bibitem [{\citenamefont {Huang}\ \emph {et~al.}(2018)\citenamefont {Huang},
  \citenamefont {Yan}, \citenamefont {Li}, \citenamefont {Khan}, \citenamefont
  {Zhang}, \citenamefont {Pi}, \citenamefont {Yu},\ and\ \citenamefont
  {Yang}}]{Huang18}%
  \BibitemOpen
  \bibfield  {author} {\bibinfo {author} {\bibfnamefont {K.}~\bibnamefont
  {Huang}}, \bibinfo {author} {\bibfnamefont {Y.}~\bibnamefont {Yan}}, \bibinfo
  {author} {\bibfnamefont {K.}~\bibnamefont {Li}}, \bibinfo {author}
  {\bibfnamefont {A.}~\bibnamefont {Khan}}, \bibinfo {author} {\bibfnamefont
  {H.}~\bibnamefont {Zhang}}, \bibinfo {author} {\bibfnamefont
  {X.}~\bibnamefont {Pi}}, \bibinfo {author} {\bibfnamefont {X.}~\bibnamefont
  {Yu}}, \ and\ \bibinfo {author} {\bibfnamefont {D.}~\bibnamefont {Yang}},\
  }\bibfield  {title} {\enquote {\bibinfo {title} {High and fast response of a
  graphene–silicon photodetector coupled with 2d fractal platinum
  nanoparticles},}\ }\href {\doibase 10.1002/adom.201700793} {\bibfield
  {journal} {\bibinfo  {journal} {Advanced Optical Materials}\ }\textbf
  {\bibinfo {volume} {6}},\ \bibinfo {pages} {1700793} (\bibinfo {year}
  {2018})},\ \Eprint
  {http://arxiv.org/abs/https://onlinelibrary.wiley.com/doi/pdf/10.1002/adom.201700793}
  {https://onlinelibrary.wiley.com/doi/pdf/10.1002/adom.201700793} \BibitemShut
  {NoStop}%
\bibitem [{\citenamefont {Dou}\ \emph {et~al.}(2014)\citenamefont {Dou},
  \citenamefont {Yang}, \citenamefont {You}, \citenamefont {Hong},
  \citenamefont {Chang}, \citenamefont {Li},\ and\ \citenamefont
  {Yang}}]{Dou2014}%
  \BibitemOpen
  \bibfield  {author} {\bibinfo {author} {\bibfnamefont {L.}~\bibnamefont
  {Dou}}, \bibinfo {author} {\bibfnamefont {Y.~M.}\ \bibnamefont {Yang}},
  \bibinfo {author} {\bibfnamefont {J.}~\bibnamefont {You}}, \bibinfo {author}
  {\bibfnamefont {Z.}~\bibnamefont {Hong}}, \bibinfo {author} {\bibfnamefont
  {W.-H.}\ \bibnamefont {Chang}}, \bibinfo {author} {\bibfnamefont
  {G.}~\bibnamefont {Li}}, \ and\ \bibinfo {author} {\bibfnamefont
  {Y.}~\bibnamefont {Yang}},\ }\bibfield  {title} {\enquote {\bibinfo {title}
  {Solution-processed hybrid perovskite photodetectors with high
  detectivity},}\ }\href {\doibase 10.1038/ncomms6404} {\bibfield  {journal}
  {\bibinfo  {journal} {Nature Communications}\ }\textbf {\bibinfo {volume}
  {5}},\ \bibinfo {pages} {5404 EP --} (\bibinfo {year} {2014})}\BibitemShut
  {NoStop}%
\bibitem [{\citenamefont {Konstantatos}(2018)}]{Konstantatos2018}%
  \BibitemOpen
  \bibfield  {author} {\bibinfo {author} {\bibfnamefont {G.}~\bibnamefont
  {Konstantatos}},\ }\bibfield  {title} {\enquote {\bibinfo {title} {Current
  status and technological prospect of photodetectors based on two-dimensional
  materials},}\ }\href {\doibase 10.1038/s41467-018-07643-7} {\bibfield
  {journal} {\bibinfo  {journal} {Nature Communications}\ }\textbf {\bibinfo
  {volume} {9}},\ \bibinfo {pages} {5266} (\bibinfo {year} {2018})}\BibitemShut
  {NoStop}%
\end{thebibliography}%
\end{document}